\documentclass{nature} % preprint

\usepackage{graphicx}
\usepackage{epsfig}
\usepackage{amssymb}
\usepackage{amsmath}
\usepackage{mathrsfs}
\usepackage{color}
\usepackage[square,super,sort&compress]{natbib}
\usepackage{multirow}
\usepackage{hyperref}
\hypersetup{
	colorlinks=true,
	linkcolor=blue,
	filecolor=blue,      
	urlcolor=blue,
	citecolor=blue,
}

\newcommand{\ergs}{${\rm erg \ cm^{-2} \ s^{-1}}$ }

\newcommand{\todo}{\ifmmode {\Huge \bullet} \else {\Huge$\bullet$}\fi}

\newcommand{\kms}{\ifmmode {\rm km\,s}^{-1} \else km\,s$^{-1}$ \fi}

\newcommand{\ergcms}{\ifmmode {\rm ergs\,cm}^{-2}\,{\rm s}^{-1} \else ergs\,cm$^{-2}$\,s$^{-1}$\fi}
\newcommand{\ergcmsA}{\ifmmode{\rm ergs}\, {\rm cm}^{-2}\,{\rm s}^{-1}\,{\rm\AA}^{-1} \else ergs\, cm$^{-2}$\, s$^{-1}$\, \AA$^{-1}$\fi}
\newcommand{\ergcmsHz}{\ifmmode{\rm ergs\,cm}^{-2}\,{\rm s}^{-1}\,{\rm Hz}^{-1} \else ergs\,cm$^{-2}$\,s$^{-1}$\,Hz$^{-1}$\fi}
\newcommand{\phcms}{\ifmmode {\rm ph\,cm}^{-2}\,{\rm s}^{-1} \else ,ph\,cm$^{-2}$\,s$^{-1}$\fi}
\newcommand{\phcmsA}{\ifmmode {\rm ph\,cm}^{-2}\,{\rm s}^{-1}\,{\rm\AA}^{-1} \else ph\,cm$^{-2}$\,s$^{-1}$\,\AA$^{-1}$\fi}

\newcommand\Msun{\ifmmode M_{\odot} \else $M_{\odot}$\fi}
\newcommand\msun{\ifmmode M_{\odot} \else $M_{\odot}$\fi}
\newcommand\Lsun{\ifmmode L_{\odot} \else $L_{\odot}$\fi}
\newcommand\Zsun{\ifmmode Z_{\odot} \else $Z_{\odot}$\fi}
\newcommand\mpyr{\ifmmode \Msun\,{\rm yr}^{-1} \else $\Msun\,{\rm yr}^{-1}$ \fi}

\newcommand{\Luv}{\ifmmode L_{1450} \else $L_{1450}$\fi}
\newcommand{\Lop}{\ifmmode L_{5100} \else $L_{5100}$\fi}
\newcommand{\Lthree}{\ifmmode L_{3000} \else $L_{3000}$\fi}
\newcommand{\lledd}{\ifmmode L/L_{\rm Edd} \else $L/L_{\rm Edd}$\fi}
\newcommand{\ledd}{\ifmmode L_{\rm Edd} \else $L_{\rm Edd}$\fi}
\newcommand{\lamLlam}{\ifmmode \lambda L_{\lambda} \else $\lambda L_{\lambda}$\fi}
\newcommand{\lbol} {\ifmmode L_{\rm bol} \else $L_{\rm bol}$\fi}
\newcommand{\llbol}{\ifmmode \log\left(\lbol/\ergs\right) \else $\log\left(\lbol/\ergs\right)$\fi}

\newcommand{\fuv}{\ifmmode f_{\lambda}\left(1450\AA\right) \else $f_{\lambda}\left(1450 {\rm \AA}\right)$\fi}
\newcommand{\fthree}{\ifmmode f_{\lambda}\left(3000\AA\right) \else $f_{\lambda}\left(3000{\rm \AA}\right)$\fi}
\newcommand{\fH}{\ifmmode f_{\lambda}\left(1.65\micron\right) \else
$f_{\lambda}\left(1.65\micron\right)$\fi}

\newcommand{\mbh}{\ifmmode M_{\rm BH} \else $M_{\rm BH}$\fi}
\newcommand{\lmbh}{\ifmmode \log\left(\mbh/\Msun\right) \else $\log\left(\mbh/\Msun\right)$\fi}

\newcommand \Hbeta {\ifmmode {\rm H}\beta \else H$\beta$\fi}
\newcommand \hb    {\ifmmode {\rm H}\beta \else H$\beta$\fi}
\newcommand  \mgii  {\ifmmode {\rm Mg}{\textsc{ii}} \else Mg\,{\sc ii}\fi}
\newcommand  \MGII  {\ifmmode {\rm Mg}\,{\sc ii}\,\lambda2798 \else Mg\,{\sc ii}\,$\lambda2798$\fi}
\newcommand  \siiv  {\ifmmode {\rm Si}\, {\sc iv}\ \else Si\,{\sc iv}\fi}
\newcommand  \SIIV  {\ifmmode {\rm Si}\,{\sc iv}\,\lambda1399 \else Si\,{\sc iv}\,$\lambda1399$\fi}
\newcommand  \civ  {\ifmmode {\rm C}\, {\textsc IV}\ \else C\,{\sc IV}\fi}
\newcommand  \CIV  {\ifmmode {\rm C}\,{\sc iv}\,\lambda1549 \else C\,{\sc iv}\,$\lambda1549$\fi}
\newcommand  \NV  {\ifmmode {\rm N}\,{\sc v}\,\lambda1240 \else N\,{\sc v}\,$\lambda1240$\fi}
\newcommand  \nv  {\ifmmode {\rm N}\,{\sc v}\ \else N\,{\sc v}\fi}
\newcommand  \cv  {\ifmmode {\rm C}\,{\sc v}\ \else C\,{\sc v}\fi}
\newcommand  \LyA  {\ifmmode {\rm Ly}\,{\sc $\alpha$}\,\lambda1216 \else Ly\,{\sc $\alpha$}\,$\lambda1216$\fi}
\newcommand  \lya {\ifmmode {\rm Ly}\,{\sc $\alpha$}\ \else Ly\,{\sc $\alpha$}\fi}
\newcommand  \feii {\ifmmode {\rm Fe}\,{\textsc{ii}}\, \else Fe\,{\sc ii}\fi}

\newcommand  \aliii  {\ifmmode {\rm Al}{\textsc{iii}} \else Al\,{\sc iii}\fi}
\newcommand  \ALIII  {\ifmmode {\rm Al}\,{\sc iii]}\,\lambda1854 \else C\,{\sc iii]}\,$\lambda1854$\fi}

\newcommand  \CIII  {\ifmmode {\rm C}\,{\sc iii]}\,\lambda1909 \else C\,{\sc iii]}\,$\lambda1909$\fi}
\newcommand  \oi    {\ifmmode \left[{\rm O}\,{\textsc i}\right] \else [O\,{\sc i}]\fi}
\newcommand  \OI    {\ifmmode \left[{\rm O}\,{\textsc i}\right]\,\lambda6300 \else [O\,{\sc i}]$\,\lambda6300$ \fi}

\newcommand  \oii   {\ifmmode \left[{\rm O}\,{\textsc {ii}}\right] \else [O\,{\sc ii}]\fi}
\newcommand  \OII   {\ifmmode \left[{\rm O}\,{\textsc ii}\right]\,\lambda3729 \else [O\,{\sc ii}]\,$\lambda3729$ \fi}
\newcommand  \oiii  {\ifmmode \left[{\rm O}\,{\textsc iii}\right] \else [O\,{\sc iii}]\fi}
\newcommand  \OIII  {\ifmmode \left[{\rm O}\,{\textsc iii}\right]\,\lambda5007 \else [O\,{\sc iii}]\,$\lambda5007$\fi}

\newcommand{\lmg}{\ifmmode L\left(\mgii\right) \else $L\left(\mgii\right)$\fi}
\newcommand{\fwmg}{\ifmmode {\rm FWHM}\left(\mgii\right) \else FWHM(\mgii)\fi}
\newcommand{\fwciv}{\ifmmode {\rm FWHM}\left(\civ\right) \else FWHM(\civ)\fi}
\newcommand{\fwhm}{\ifmmode {\rm FWHM} \else FWHM\fi}

\newcommand  \NEV   {\ifmmode \left[{\rm Ne}\,{\textsc v}\right]\,\lambda3426 \else [Ne\,{\sc v}]\,$\lambda3426$ \fi}
\newcommand  \nev  {\ifmmode \left[{\rm Ne}\,{\textsc v}\right] \else [Ne\,{\sc v}]\fi}

\def\ergs{\rm erg~s^{-1}}

\def\feii{Fe {\sc ii}}
\def\oiii{[O~{\sc iii}]}

\def\ergs{${\rm erg\,s^{-1}}$}

\begin{document}

\title{Evidence for the connection between star formation rate and evolutionary phases of quasars}
\author{Zhifu Chen$^{1}$,
Zhicheng He$^{2,3}$\thanks{E-mail: zcho@ustc.edu.cn},
Luis C. Ho$^{4,5}$,
Qiusheng Gu$^{6}$,
Tinggui Wang$^{2,3}$,
Mingyang Zhuang$^{4,5}$,
Guilin Liu$^{2,3}$,
Zhiwen Wang$^1$
}
\maketitle

\begin{affiliations}
\item{Department of Physics, Guangxi University for Nationalities, Nanning 530006, China}

\item{CAS Key Laboratory for Research in Galaxies and Cosmology, Department of Astronomy, University of Science and Technology of China, Hefei, Anhui 230026, China}

\item{School of Astronomy and Space Science, University of Science and Technology of China, Hefei 230026, China}

\item{Kavli Institute for Astronomy and Astrophysics, Peking University, Beijing 100871, China}

\item{Department of Astronomy, School of Physics, Peking University, Beijing 100871, China }

\item{School of Astronomy and Space Science, Nanjing University, Nanjing 210093, China}

\end{affiliations}

\begin{abstract}
Both theory and observations suggest that outflows driven by an active central supermassive black hole (SMBH) have a feedback effect on shaping the global properties of the host galaxy\cite{Silk1998,2000MNRAS.311..576K,2005Natur.433..604D,2006ApJS..163....1H,fabian2012,zubovas2012,2014ARA&A..52..589H,he2019}. However, whether feedback from the outflow is effective, and if so, whether it is positive or negative, has long been controversial. Here, using the latest catalog from the Sloan Digital Sky Survey (SDSS), we use the flux ratio of the \oii\ to \nev\ emission lines as a proxy to compare the star formation rate (SFR) in the hosts of quasars with different types of broad absorption lines (BALs): low-ionization (Lo)BAL, high-ionization (Hi)BAL, and non-BAL. We find that SFR decreases from LoBAL to HiBAL quasars, and then increases from HiBAL to non-BAL quasars. Assuming that the sequence of LoBAL to HiBAL to non-BAL represents evolution, our results are consistent with a quenching and subsequent rebound of star formation in quasar host galaxies.  
This phenomenon can be explained that the SFR is suppressed by the outflow, which then rebounds once the outflow disappears as the quasars evolve from HiBALs to non-BALs. Our result suggests that the quasar outflow has a negative global feedback on galaxy evolution. 
\end{abstract}

Quasar outflows are primarily detected via blueshifted BALs. About $10\%-40\%$ of optically selected quasars host BALs
\cite{2009ApJ...692..758G,2011MNRAS.410..860A}. BALs are primarily detected in high-ionization lines, such as \SIIV\ and \CIV, which are usually called HiBALs. Furthermore, about 15\% of BALs are LoBALs, which exhibit blueshifted absorptions in low-ionization lines such as \MGII\ and \ALIII. There are primarily two scenarios to account for the presence or absence of BALs, or more generally, the manifestation of different subclasses of BALs.  In the orientation-dependent scenario, the diversity of the BAL phenomenon can be attributed to the different viewing angles of the same intrinsic population\cite{urry1995}. In the orientation model, there should be no statistical difference in the properties of the host galaxies of different BAL types. By contrast, the evolutinary scenario\cite{1992ApJ...399L..15B,hall2002,wang2016} suggests that different types of BAL quasars are at different stages of evolution. Ref.\cite{wang2016} estimate that outflows have a minimum covering factor of a few to 20\% for the \feii\ scattering region, which is much larger than the fraction of FeLoBAL (LoBAL exhibited by Fe lines) quasars ($<1\%$ of all quasars\cite{trump2006,2012ApJ...757..180D}). This result cannot be explained by the orientation-dependent effect alone and suggests that LoBAL quasars are at a special stage of quasar evolution.

In the evolution scenario, LoBAL quasars likely live in the early stage of quasar evolution, which, given their detection rate $<1\%$, constitutes a short transition phase from the merger-induced to the normal stage\cite{1992ApJ...399L..15B,hall2002,wang2016}. In the starburst or post-starburst phase in which these quasars live, there is enough dense gas and dust to sustain the low ionization state of LoBAL quasars. LoBAL quasars are expected to have reddened spectra, as observed\cite{2010ApJ...714..367Z}. With the consumption of gas and dust, the quasar spectra turn blue, and the outflows gradually transition from the LoBAL to the HiBAL phase. From the perspective of outflow evolution, quasars evolve along the sequence LoBALs $\rightarrow$ HiBALs $\rightarrow$ non-BALs. Comparing the SFR of these three stages is helpful to understand the feedback effect of quasar-driven outflows on galaxy evolution.

\begin{figure*}
\centering
\includegraphics[width=0.85\textwidth]{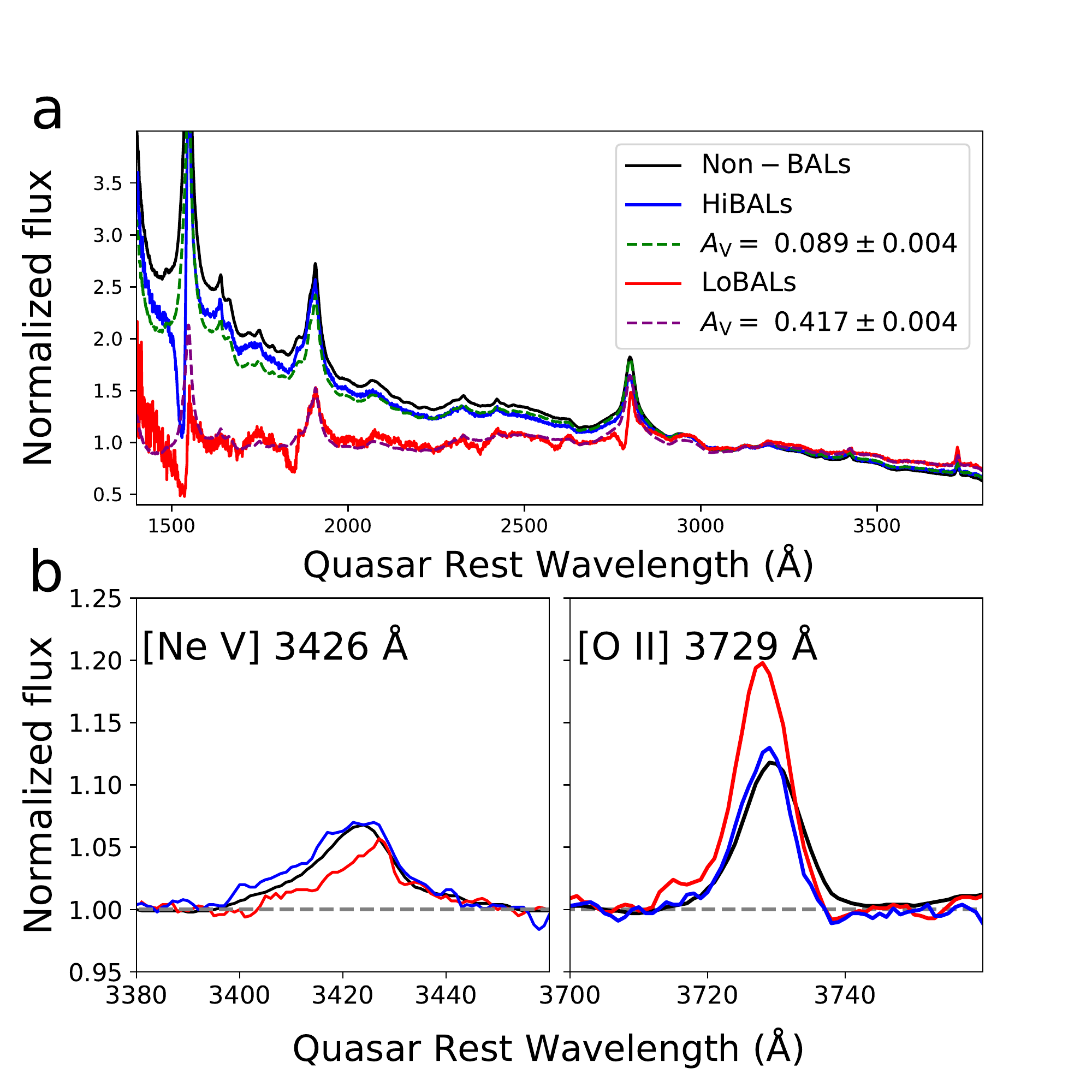}
\caption{\textbf{The median composite spectra of quasars from SDSS DR16.} 
\textbf{a}, Both the LoBALs and HiBALs are redder than the non-BALs. Relative to non-BALs, HiBALs and LoBALs have an extinction of 
$A_V = 0.089\pm0.004$ (green dashed line), and $0.417\pm0.004$ (purple dashed line) mag, respectively. 
\textbf{b}, The \nev\ and \oii\ emission lines, normalized to the local continuum in the median composite spectra.}
\label{Fig1}
\end{figure*}

Here we collect 407 LoBAL, 714 HiBAL, and $\rm14,144$ non-BAL quasars from the Sixteenth Data Release (DR16Q) of the SDSS quasar catalog\cite{2020ApJS..250....8L}. From the median composite spectra (Figure \ref{Fig1}\textbf{a}) of these three types of quasars (see Methods for details), it is clearly seen that LoBALs are much redder than both HiBALs and non-BALs, and that HiBALs are slightly redder than non-BALs. If we invoke the extinction curve of the Small Magellanic Cloud, $A_{\lambda}=1.39\lambda^{-1.2}E(B-V)$\cite{2006MNRAS.367..945Y}, where $E(B-V)$ is the color excess, to describe the dust reddening within BAL quasars relative to non-BAL quasars, we obtain $A_V=0.417\pm0.004$ mag and $A_V=0.089\pm0.004$ mag for LoBAL and HiBAL quasars, respectively. The relative differences in reddening among LoBALs, HiBALs, and non-BALs is consistent with previous studies\cite{2009ApJ...692..758G,2010ApJ...714..367Z}. The evolution scenario explains the reddening 
differences in terms of the expulsion of a thick shroud of gas and dust during the BAL phase. 

\begin{figure*}
\centering
\includegraphics[width=12cm]{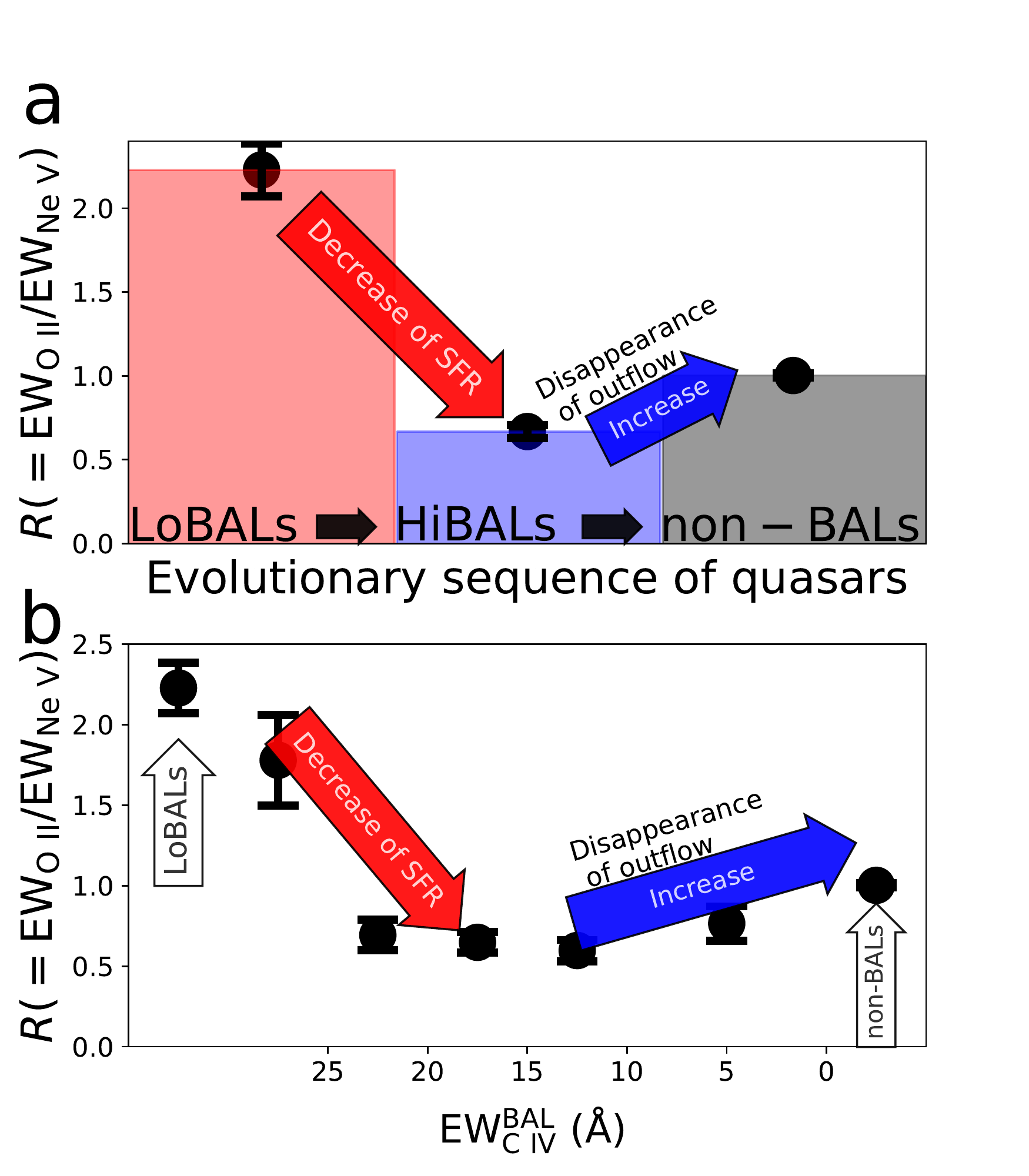}
\caption{\textbf{Adopting the line ratio $R = {\rm EW_{\oii}/EW_{\nev}}$ as a proxy to investigate the star formation at 
different phases of quasar evolution.}
\textbf{a}, The line ratio significantly decreases and then increases as quasars evolve from LoBALs $\rightarrow$ HiBALs $\rightarrow$ non-BALs. 
The vertical error bars mark the 1$\sigma$ uncertainty of the line ratio.
\textbf{b}, The variation of $R$ as a function of the \civ\ broad absorption line strength ${\rm EW_{\civ}^{BAL}}$. The level of star formation activity, as reflected in $R$, increases as ${\rm EW_{\civ}^{BAL}}$ decreases.}
\label{Fig2}
\end{figure*}

The narrow-line regions and extended emission-line regions in quasars can produce both the low-ionization \OII\ and the high-ionization \NEV\ emission. Young massive stars within quasar host galaxies can easily excite \OII\ emission\cite{ho2005}, but not \NEV. Therefore, the star formation activity of quasar hosts can be gauged by the relative strength of the \OII\ and \NEV\ emission lines\cite{2018MNRAS.480.5203M}. 
A larger ratio of $R={\rm EW}_{\oii}/{\rm EW}_{\nev}$ indicates a higher SFR.  Figure \ref{Fig1}\textbf{b} highlights the \nev\ and \oii\ emission lines, normalized to the underlying continua (see Methods), of the composite spectra of the three classes of quasars.  As listed in Table \ref{Tab1} and graphically shown in Figure \ref{Fig2}\textbf{a}, $R = 2.228\pm0.158$, $0.668\pm0.039$, and $1.002\pm0.013$ for the LoBAL, HiBAL, and non-BAL quasars, respectively.  Defining the significance of the difference of $R$ between two different BAL quasar types as $N_{\sigma}=|R_1-R_2|/\sqrt{\sigma_{R_1}^2+\sigma_{R_2}^2}$, we can clearly see that $R$ markedly ($9.6\,\sigma$) decreases to a much lower value and then significantly ($8.1\,\sigma$) increases, as the quasars evolve from LoBALs $\rightarrow$ HiBALs $\rightarrow$ non-BALs. To further confirm the details of this rebound phenomenon, we divide the HiBAL quasars into five bins according to a sequence of decreasing \civ\ BAL strength: ${\rm EW_{\civ}^{BAL}} >25$, $20-25$, $15-20$, $10-15$, and $<10$ \AA. The corresponding values of $R$ are $1.779\pm0.282$, $0.693\pm0.095$, $0.648\pm0.064$, $0.597\pm0.066$, and $0.765\pm0.106$.  As shown in \ref{Fig2}\textbf{b}, $R$ also shows a rebound phenomenon as ${\rm EW_{\civ}^{BAL}}$ decreases, suggesting that the SFR rebound phenomenon may begin during the HiBAL phase, when ${\rm EW_{\civ}^{BAL}}<10$ \AA. Although we do not have an absolute measurement of the SFR, the above results suggest that star formation within host galaxies first drops and then bounces back as quasars evolve from LoBALs $\rightarrow$ HiBALs $\rightarrow$ non-BALs. Some studies\cite{Willott2003,Priddey2007} find that the far-IR emission of quasar host galaxies does not depend on the presence of BALs. One reason may be that their samples do not include LoBALs or BALs with large ${\rm EW_{\civ}^{BAL}}$. A correlation between SFR and ${\rm EW_{\civ}^{BAL}}$ has been reported\cite{Priddey2007}, but the the small sample size (15 objects) precludes a definitive conclusion ($1\%-4\%$ significance from a Kolmogorov–Smirnov test).

The interpretation of our results may be complicated by the intrinsic diversity of narrow-line emission in quasars. For example, LoBAL quasars tend to show weak \oiii\ emission\cite{boroson1992}, and \oiii-weak LoBALs on average have undetected \nev\cite{zhang2010}. Is the large value of $R$ in LoBALs caused by weak \nev\ or strong \oii?  
Although the average EW of \nev\ in LoBALs ($0.808\pm 0.055$ \AA) is systematically lower than that of HiBALs ($1.667\pm 0.036$ \AA), the average EW of \oii\ changes by an large enough amount: $1.789\pm 0.032$ \AA\ for LoBALs compared to $1.113\pm 0.061$ \AA\ for HiBALs. As a result, $R$ is still significantly ($\sim 10\ \sigma$) larger in LoBALs ($1.073\pm 0.031$) than in HiBALs ($0.668\pm 0.039$) even after we use the \nev\ EW of HiBALs to calculate $R$ value for LoBALs. All three classes have similar spectral energy distribution (SED; 
Extended Fig. 1), and both $R$ and the differences in $R$ among the three classes are independent of black hole mass (Table \ref{Tab2} and Extended Fig. 2). Photoionization calculations\cite{2019ApJ...882...89Z} predict that solar metallicity, radiation pressure-dominated narrow-line regions should have an intrinsic ratio $R \approx 0.5$.  Therefore, any excess above this baseline value can be regarded as \oii\ emission arising from star formation.  Since the three classes of quasars considered here have $R \gg 0.5$, all three populations---especially LoBALs and non-BALs---have substantial levels of ongoing star formation.  Since a fraction of BAL quasars that are not along our line-of-sight will be misidentified as non-BAL quasars, the actual value of $R$ of non-BALs, and hence their degree of star formation rebound, should actually be higher than what is observed.  The possible contribution from shocks poses a complication, for models of shocks with a precursor\cite{2008ApJS..178...20A} that assume solar abundances, preshock particle density $1~\rm cm^{-3}$, and velocities $100-1000$ \kms\ can generate $R > 10$.  The ratios $R$ for all three populations (LoBAL, HiBAL, and non-BAL quasars) are well below those of shock-dominated regions ($>10$), indicating that \oii\ and \nev\ lines are not excited by shocks. Moreover, if shocks from BAL outflows were an important contributor to the narrow-line excitation, we would expect HiBALs to have larger $R$ than non-BALs, the opposite of what is observed.  

The SFR rebound phenomenon reflects the important role that quasar-driven outflows play in galaxy evolution. We propose that during the early stage of quasar evolution, LoBALs have enough cold gas to sustain intense star formation, and their outflow has not yet expanded to the outer star formation region. Our sample contains 64 LoBAL, 714 HiBAL, and 14,144 non-BAL quasars in the redshift range $1.4 < z < 1.6$. From the perspective of an evolutionary scenario, the LoBAL and HiBAL phases occupy, respectively, 0.43\% and 4.8\% of the total quasar lifetime. If we suppose a quasar lifetime of 50 Myr\cite{hopkins2005}, then the LoBAL phase lasts 0.2 Myr. 
However, it should be noted that LoBALs are often much redder than other quasars. As a result, the fraction of LoBALs is underestimated in optically selected samples like SDSS\cite{urrutia2009}.
For a typical BAL outflow velocity of 10,000 \kms, a freely expanding outflow can reach as far as 2 kpc from the center of the galaxy during the LoBAL phase. If the outflow couples to the interstellar medium, the shock velocity will be much slower, the galactocentric distance impacted much smaller, and significant star formation can occur prior to the ``blow-out" phase\cite{hopkins2006}.
Analytical models\cite{Silk1998,zubovas2012} show that an energetic, high-velocity wind that carries a few percent of a quasar's luminosity can couple to the interstellar medium and drive galaxy-wide outflows, which can effectively sweep out the fuel for star formation or prevent gas cooling. IRAS F11119+3257\cite{tombesi2015} and Mrk 231\cite{feruglio2015} are examples where the molecular outflows have been found to be driven by an inner, energy-conserving ultrafast outflow\cite{laha2021}. As a result, the SFR begins to decline dramatically when the quasars evolve from the LoBAL to the HiBAL phase. Considering a HiBAL fraction of 5\% and a quasar lifetime of 50 Myr\cite{hopkins2005}, the duration of the HiBAL phase is $\sim$ 2.5 Myr. We should note that we select BALs using standard convention, namely that they have line widths $>$ 2000 \kms at depths $>$ 10\% below the continuum and then check visually. As a result, the fraction of BALs is lower than the typical 
value 10-40\% in previous studies. If we take the BAL fraction value 10-40\%, the duration of the HiBAL phase can be up to ten Myr. Therefore, the outflow has plenty of time to reach and thereby potentially impact galactic scales. Some studies\cite{arav2018} find that 50\% of BAL outflows are located at distances larger than 100 pc, and at least 12\% are on scales $>1$ kpc.  Since the ionization energy of O$^+$ (13.6 eV) coincides closely to the peak effective temperature of O-type stars, and the lifetime of young, massive ionizing stars (a few Myr) is also comparable to the duration of BAL outflows, we can witness the decrease of \oii\ flux during the HiBAL phase. The suppression of star formation terminates when the outflow subsides as the quasar evolves from the HiBAL to the non-BAL phase, and star formation rekindles. 
Our discovery of SFR variation and its close connection to the BAL phenomenon strongly suggests that quasar outflows impart a negative feedback on global, galactic scale. At the same time, the influence evidently is transitory and closely linked to the life cycle of quasar activity.

%\clearpage
%\newpage
\begin{methods}
\subsection{Quasar Samples.}
The SDSS gathered spectra in the wavelength range $3800-9200$ \AA\ at a resolution of $R\approx2000$ from 2000 to 2008 (SDSS-I/II)\cite{2009ApJS..182..543A}, and in the wavelength range $3600-\rm10~500$ \AA\ at a resolution of $R\approx1300-2500$ from 2008 to 2020 (SDSS-III/IV)\cite{2013AJ....146...32S,2013AJ....145...10D,2016AJ....151...44D}. DR16Q is the final dataset for the SDSS-IV quasar catalog, which contains 750,414 quasars accumulated from SDSS-I to SDSS-IV\cite{2020ApJS..250....8L}. We use the spectroscopic data from DR16Q to construct samples of LoBALs, HiBALs, and non-BALs. In order to investigate the star formation activity of quasar host galaxies with \OII\ emission, all the spectra should cover rest-frame $\lambda<4000$ \AA. For each spectrum, we fit a power-law continuum ($f_{\lambda}=A\lambda^{\alpha}$) plus an iron template \cite{2001ApJS..134....1V,2004A&A...417..515V} in line-free regions that are not obviously contaminated by strong emission or absorption lines. We use multiple Gaussians to model the \CIV\ and \MGII\ emission lines. We define BALs using standard convention, namely that they have line widths $>2000$ \kms\ at depths $>10\%$ below the continuum \cite{1991ApJ...373...23W}.

\textbf{LoBAL quasar sample.} We require the LoBALs that include \MGII\ BALs. Accounting for the wavelength coverage of the SDSS spectra, we search for LoBALs in the range $0.38<z_{\rm em}<1.6$. We obtain 407 quasars with robust \mgii\ BALs.

\textbf{HiBAL quasar sample.} We require the HiBAL quasars that include \CIV\ BALs. Accounting for the wavelength coverage of the SDSS spectra,, we search for HiBALs in the range $1.4<z_{\rm em}<1.6$. We obtain 714 quasars with \civ\ BALs, but without \aliii, \feii, and \mgii\ BALs.

\textbf{Non-BAL quasar sample.} The non-BAL sample contains quasars in the range $1.4<z_{\rm em}<1.6$. This redshift range guarantees quasar spectra that do not produce \civ, \aliii, \feii, and \mgii\ BALs. In order to reliably characterize the quasar continuum, we only consider the spectra with median signal-to-noise ratio $\ge5$. This results in $\rm14,144$ robust quasars without BALs.

For the quasars included in LoBAL, HiBAL, and non-BAL samples, we determine their redshifts from the narrow \nev, \oii, or \oiii\ emission lines, when available; otherwise, we directly adopt the redshifts given in the DR16Q quasar catalog\cite{2020ApJS..250....8L}.

\subsection{Composite spectra.}\label{sect:Composite_spectra}
We create median composite spectra for the LoBAL, HiBAL, and non-BAL quasars, following the method used by previous works \cite{2012ApJ...748..131S,2020ApJ...893...25C}. These composite spectra are shown in Figure \ref{Fig1}, which have been normalized to the emission at 3000 \AA. It is clearly seen from Figure \ref{Fig1} that the composite spectra of both LoBAL and HiBAL quasars are significantly redder than that of non-BAL quasars, which can be ascribed to dust extinction within LoBAL and HiBAL quasars. We adopt the Small Magellanic Cloud extinction curve, $A_{\lambda}=1.39\lambda^{-1.2}E(B-V)$\cite{1984A&A...132..389P,1992ApJ...395..130P,2006MNRAS.367..945Y}. The results are shown with green (HiBALs) and purple (LoBALs) dot-dashed lines in Figure \ref{Fig1}.

To measure the \NEV\ and \OII\ emission lines in the composite spectra,
we fit a power-law continuum plus an iron template \cite{2001ApJS..134....1V,2004A&A...417..515V} in line-free regions.  
The iron template is fit with four free parameters: velocity dispersion, velocity shift, and separate amplitudes for the regions 2200--3100 and 3250--3500 \AA, while avoiding 3100--3250 \AA. This piece-meal approach was found necessary given the current incomplete knowledge of the iron spectrum over this challenging wavelength region.
The fitting results are shown in Supplementary Figures 1--3.
The emission-line strengths are measured in the composite spectra normalized by the fits of the power-law plus iron continuum.
The equivalent widths of \nev\ and \oii\ are obtained by integrating the emission line profiles in the normalized composite spectra. The results are listed in Table \ref{Tab1}. We repeat the above fitting process 1000 times, each time adding a Gaussian random error to the observed flux. The mean and standard deviation of these fitting trials give the final result and corresponding error.

The composite spectra are stacked in the quasar rest frame. The uncertainty on the quasar redshift might smear out the weak narrow lines of \nev\ and \oii\ during the stacking process, and contribute error to the strength ratio $R = {\rm EW_{\oii}/EW_{\nev}}$. In order to estimate the error on $R$ due to the uncertainty of quasar redshift, we perturb the redshift by $\pm500$ \kms, which represents a realistic estimate of the true redshift uncertainty. Then we reproduce the composite spectra and redo the line measurements.  We find that the standard deviations of $R$ are 0.112, 0.019, and 0.001 for LoBALs, HiBALs, and non-BALs, respectively.  These values are smaller than the errors contributed from the flux uncertainties (the last column of Table \ref{Tab1}).

\subsection{Possible selection effects due to redshift mismatch.} The redshift range of the LoBAL sample ($0.38<z<1.6$) is wider than that of the HiBAL and non-BAL samples ($1.4<z<1.6$).  To investigate the potential effect of redshift mismatch, we isolate the subset of 64 LoBALs whose redshift range ($1.4<z<1.6$) matches that of the HiBAL and non-BAL samples. The composite spectrum of this matched subsample of LoBALs yields $R = 2.317\pm0.769$, which is entirely consistent with $R=2.228\pm0.158$ for the original sample.  We therefore conclude that the observed differences in $R$ among the quasar classes are not due to possible selection effects due to redshift mismatch.

\subsection{Possible influence of dust-reddened quasars in the non-BAL sample.} A portion of non-BAL quasars have a redder color, which might be due to dust reddening. As dust-reddened quasars exhibit enhanced \oii\ emissions with respect to unreddened quasars\cite{Richards2003}, they might result in higher values of $R$. The distribution of the spectral indices of non-BAL quasars ($f_{\lambda}\propto\lambda^{\alpha}$) measured between 1600 \AA\ and 3000 \AA\ reveals an extended tail with $\alpha>-0.85$ (Supplementary Figure 4).  Excluding this tail of reddened objects, which comprises $\sim 10\%$ of the sample, results in $R = 1.051\pm0.013$, which is indistinguishable from the full sample ($R = 1.002\pm0.013$).  We therefore conclude that the possible presence of dust-reddened quasars among non-BALs does not affect our results.

\subsection{Possible influence of the Baldwin effect.} The Baldwin effect\cite{baldwin1977}, namely, the anti-correlation between the equivalent width of emission lines and continuum luminosity, is stronger for \nev\ than for \oii\ \cite{2013ApJ...762...51Z}. The left panel of Supplementary Figure 5 shows the distributions of 3000 \AA\ luminosity for the LoBAL, HiBAL, and non-BAL quasars; the median values and standard deviations of the logarithmic luminosities are, respectively, (44.88,0.48), (45.00,0.29), and (44.97,0.21). Although the luminosity distributions are different among the three classes, they have similar median values.  We therefore conclude that the evolution trend deduced from the variation of $R$ among the three quasar classes is not affect by the Baldwin effect.

\subsection{Possible influence of the spectral energy distribution.} Apart from differences due to reddening, the spectral energy distribution (SED) of LoBALs is consistent with that of non-BALs from the UV to the mid-IR\cite{2017ApJ...848..104S}. Extended Fig. 1 confirms that after correcting for extinction, the overall continuum shapes of the composite spectra of the three classes are indeed very similar. This suggests that the differences in $R$ among the three quasar classes are unlikely to be caused by differences in their SEDs.

\subsection{Possible influence of different black hole masses.}
The SFR of a galaxy depends on its stellar mass, both for star-forming galaxies\cite{brinchmann2004,daddi2007}
and for the host galaxies of quasars \cite{xie2021}.
Since our quasar sample presumably spans a broad range of host galaxy masses, it is important to ensure that our conclusion that the three classes of quasars have different star formation properties is not an artifact of their having different host galaxy masses.  While we do not have direct access to the stellar masses of the quasar host galaxies, we can use the black hole mass as a proxy for stellar mass, in light of the empirical relation between these two quantities \cite{2013ARA&A..51..511K,greene2020}.
We estimate virial black hole masses for our sample using the line width of the \mgii\ emission line obtained from multi-component Gaussian fits \cite{2019ApJS..244...36C}. All the fits are visually inspected individually. The black hole mass follows from
\begin{equation}\label{eq:MBH}
 \log\, (\frac{M_{\rm BH}}{M_\odot}) = a + b\times \log\, (\frac{L_{3000}}{10^{44}~{\rm erg~s}^{-1}}) + 2\times \log\, (\frac{{\rm FWHM_{Mg~II}}}{{\rm km~s}^{-1}}),
\end{equation}
where the coefficients $(a,b)=(0.82,0.5)$ are empirically calibrated\cite{2019ApJS..244...36C}, ${\rm FWHM}_{\mgii}$ is the full width at half-maximum of broad \mgii\ emission line, and $L_{3000}$ is the monochromatic continuum luminosity at 3000 \AA. Supplementary Figure 5 shows the black hole mass of the LoBAL (red squares), HiBAL (blue stars), and non-BAL quasars (black circles).  Note that while the black hole mass can be well determined using the \mgii\ emission line for most HiBAL and non-BAL quasars, in LoBAL quasars the blue wing of the \mgii\ line can be seriously corrupted by absorption.  Under these circumstances only the red wing of the line can be used to fit the profile, and the resulting black hole masses for LoBALs can be quite uncertain.

We divide the quasar samples into three bins based on the black hole mass. The normalized spectra of \nev\ and \oii\ emission lines are shown in Supplementary Figure 6, and the line measurements are listed in Table \ref{Tab2} and plotted in Extended Fig. 2. 
It is clearly seen that each type of quasars has similar $R$
in each of the three bins of black hole mass. Most crucially, the evolution trend of $R$ discussed in this paper is qualitatively very similar for all three mass bins. We conclude that the principal conclusion of this paper, that quasars evolve from LoBALs to HiBALs and then finally to non-BALs, is not an artifact due to mismatched host galaxy masses.

\subsection{Composite spectra with different ${\rm EW_{\civ}^{BAL}}$.}
The \civ\ absorption strength of LoBAL quasars is generally stronger than that of HiBAL quasars. In the evolution scenario, the evolution time series of the HiBALs with only weak \civ\ absorption is similar to that of the non-BALquasarss, whereas the evolution time series of HiBALs with much stronger \civ\ absorption is closer to that LoBALs. We merge the HiBALs and LoBALs with \civ\ BALs, and then divide them into five bins based on ${\rm EW_{C~IV}}$. The composite spectra of these five bins are shown in Supplementary Figures 7, and the measurements of \nev\ and \oii\ emission are listed in Supplementary Table 1. The strength ratios $R$ are shown in Figure \ref{Fig2}\textbf{b}. Quasars with the strongest \CIV\ absorption clearly have a high value of $R$ that resembles the value of the LoBAL quasars; by contrast, quasars with the weakest \civ\ absorption have $R$ values similar to those of non-BAL quasars.  Sources with intermediate levels of \civ\ absorption have lower $R$ those sources with the strongest and weakest \civ\ absorption. In other words, $R$ first decreases rapidly, then becomes stable, and finally slowly rises back to a higher value, as the \civ\ absorption strength gradually weakens.

\subsection{Data availability}
The datasets that support the figures within this paper are attached.
The SDSS sample data for different types of quasars in this work is available at\\
\url{http://staff.ustc.edu.cn/\~zcho/sample.html}. 
Any additional data are available from the corresponding author.

\subsection{Code availability}
The codes that support the figures within this paper and other findings of this
study are available from the corresponding author.

\end{methods}

\begin{addendum}
\item[Correspondence and request for materials] should be addressed to Zhicheng He (zcho@ustc.edu.cn).

\item[Acknowledgements] Zhi-Fu Chen is supported by the National Natural Science Foundation of China (12073007), the Guangxi Natural Science Foundation (2019GXNSFFA245008; GKAD19245136; 2018GXNSFAA050001), the National Natural Science Foundation of China (11763001), and the Scientific Research Project of Guangxi University for Nationalities (2018KJQD01).  Zhicheng He is supported by NSFC-11903031 and USTC Research Funds of the Double First-Class Initiative YD 3440002001.  Luis C. Ho is supported by the National Science Foundation of China (11721303, 11991052) and the National Key R\&D Program of China (2016YFA0400702).

Funding for the Sloan Digital Sky Survey IV has been provided by the Alfred P. Sloan Foundation, the U.S. Department of Energy Office of Science, and the Participating Institutions.  SDSS-IV acknowledges support and resources from the Center for High Performance Computing  at the University of Utah. The SDSS website is www.sdss.org.  SDSS-IV is managed by the Astrophysical Research Consortium for the Participating Institutions of the SDSS Collaboration including the Brazilian Participation Group, the Carnegie Institution for Science, Carnegie Mellon University, Center for Astrophysics | Harvard \& Smithsonian, the Chilean Participation Group, the French Participation Group, Instituto de Astrof\'isica de Canarias, The Johns Hopkins University, Kavli Institute for the Physics and Mathematics of the Universe (IPMU) / University of Tokyo, the Korean Participation Group, Lawrence Berkeley National Laboratory, Leibniz Institut f\"ur Astrophysik Potsdam (AIP),  Max-Planck-Institut f\"ur Astronomie (MPIA Heidelberg), Max-Planck-Institut f\"ur Astrophysik (MPA Garching), Max-Planck-Institut f\"ur Extraterrestrische Physik (MPE), National Astronomical Observatories of China, New Mexico State University, New York University, University of Notre Dame, Observat\'ario Nacional / MCTI, The Ohio State University, Pennsylvania State University, Shanghai Astronomical Observatory, United Kingdom Participation Group, Universidad Nacional Aut\'onoma de M\'exico, University of Arizona, University of Colorado Boulder, University of Oxford, University of Portsmouth, University of Utah, University of Virginia, University of Washington, University of Wisconsin, Vanderbilt University, and Yale University.

\item[Author contributions] Z.-F. C. made the calculations, wrote the manuscript and comprehensively discussed the idea.  
Z.-C. H. presented the idea, discussed the calculations, wrote the main text of the manuscript. L. C. Ho oversaw and revised the whole manuscript. Q. -S. G, T.-G. W and M. -Y. Z discussed the idea and calculations. G. -L. L and Z. -W. W gave comments on the revision of the manuscript.
All authors discussed and gave comments on the contents of the paper.

\item[Competing Interests] The authors declare that they have no competing financial interests.
\end{addendum}
%\bibliography{ref}
%\bibliographystyle{abbrv}

\begin{table}[htbp]
\caption{The equivalent widths of \nev\ and \oii\ for different types of quasars. The error ranges are
the 1$\sigma$ uncertainties.} 
\centering \tabcolsep 2mm
\label{Tab1}
\begin{tabular}{cccccc}
\hline\hline\noalign{\smallskip}
Samples   & ${\rm EW_{\nev}}$  &${\rm EW_{\oii}}$   & $R=\frac{{\rm EW_{\oii}}}{{\rm EW_{\nev}}}$  & $\sigma$ \\
          &\AA           &\AA           &         &  &               \\
\hline\noalign{\smallskip}
LoBALs    & 0.808$\pm$0.055   & 1.789$\pm$0.032   & 2.228  &  0.158   \\
HiBALs    & 1.667$\pm$0.036   & 1.113$\pm$0.061   & 0.668  &  0.039   \\
Non-BALs  & 1.246$\pm$0.007   & 1.250$\pm$0.013   & 1.002  &  0.013  \\
\hline\hline\noalign{\smallskip}
\end{tabular}
\\
\end{table}

\begin{table}[htbp]
\caption{The equivalent widths of \NEV\ and \OII\ of the quasars with different black hole masses.
The error ranges are the 1$\sigma$ uncertainties.} 
\centering \tabcolsep 2mm
\label{Tab2}
\begin{tabular}{cccccc}
\hline\hline\noalign{\smallskip}
Samples   & ${\rm EW_{\nev}}$  &${\rm EW_{\oii}}$   & $R = \frac{{\rm EW_{\oii}}}{{\rm EW_{\nev}}}$ & $M_{\rm BH}$\\
          &(\AA)           &(\AA)           &              &         $(M_{\odot}$) \\
\hline\noalign{\smallskip}
LoBALs    & 0.946$\pm$0.145   & 1.914$\pm$0.160   & 2.024$\pm$0.353 & \multirow{3}{*}{$\le10^{8.5}$}  \\
HiBALs    & 2.547$\pm$0.209   & 1.634$\pm$0.191   & 0.642$\pm$0.092   \\
Non-BALs  & 1.308$\pm$0.044   & 1.301$\pm$0.051   & 0.995$\pm$0.052   \\
\hline\noalign{\smallskip}
LoBALs    & 0.594$\pm$0.064   & 1.081$\pm$0.086   & 1.819$\pm$0.245 & \multirow{3}{*}{$10^{8.5}-10^{9.0}$}  \\
HiBALs    & 1.579$\pm$0.061   & 1.077$\pm$0.086   & 0.682$\pm$0.061   \\
Non-BALs  & 1.295$\pm$0.013   & 1.287$\pm$0.034   & 0.994$\pm$0.028   \\
\hline\noalign{\smallskip}
LoBALs    & 0.434$\pm$0.069   & 1.042$\pm$0.120   & 2.402$\pm$0.471 & \multirow{3}{*}{$\ge10^{9.0}$}  \\
HiBALs    & 1.105$\pm$0.051   & 0.760$\pm$0.065   & 0.687$\pm$0.066   \\
Non-BALs  & 1.181$\pm$0.018   & 1.147$\pm$0.029   & 0.971$\pm$0.029   \\
\hline\hline\noalign{\smallskip}
\end{tabular}
\end{table}

\addtocounter{figure}{-2}
\begin{figure*}
\renewcommand{\figurename}{Extended Data Figure}
\centering
\includegraphics[width=0.95\textwidth]{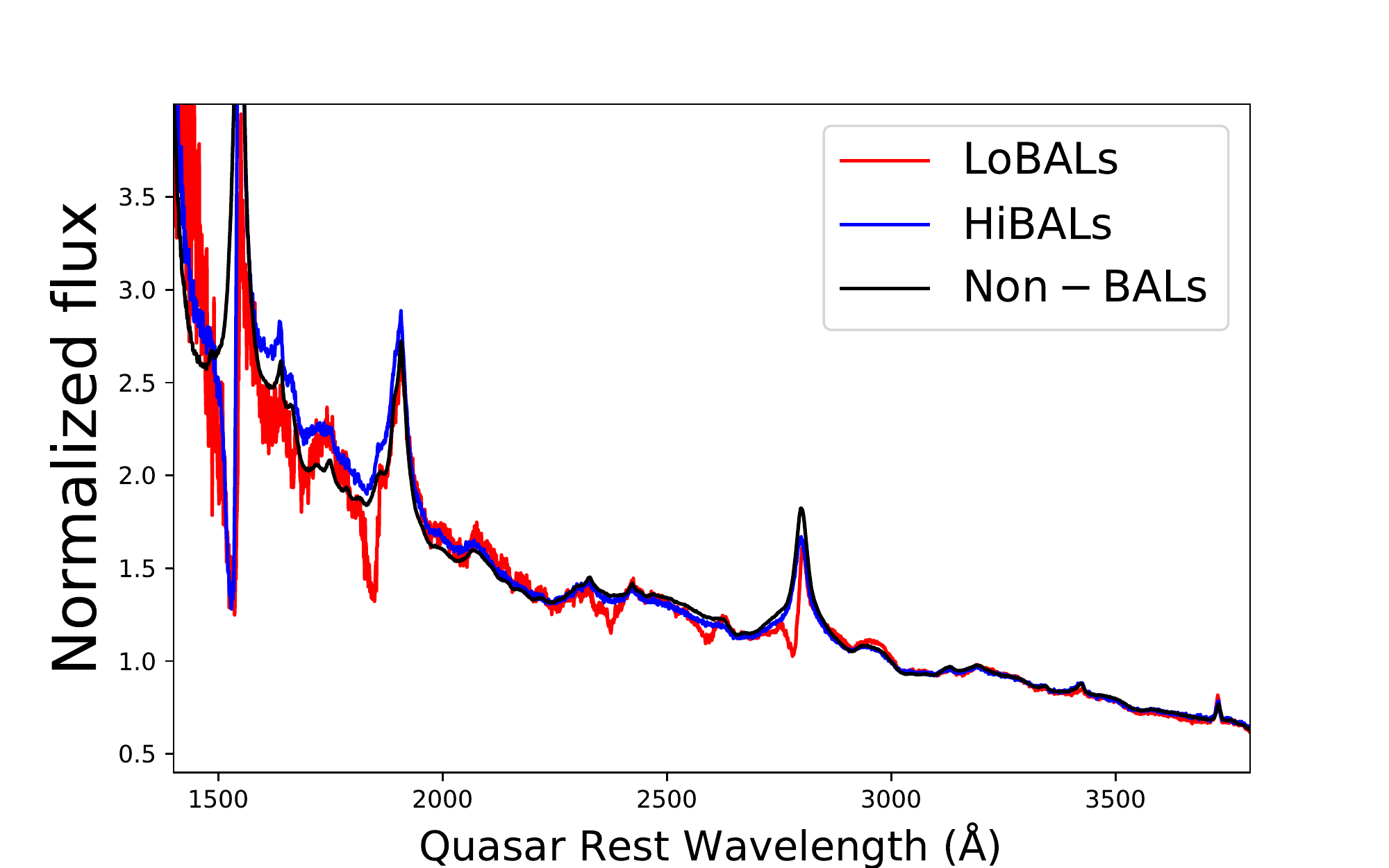}
\caption{\textbf{The composite spectra after reddening corrections.} Using the Small Magellanic Cloud extinction curve 
with $A_V = 0.088$ mag for the HiBALs (blue line) and $0.417$ mag for the LoBALs (red line).}
\label{EDFig1}
\end{figure*}

\begin{figure*}
\renewcommand{\figurename}{Extended Data Figure}
\centering
\includegraphics[width=0.99\textwidth]{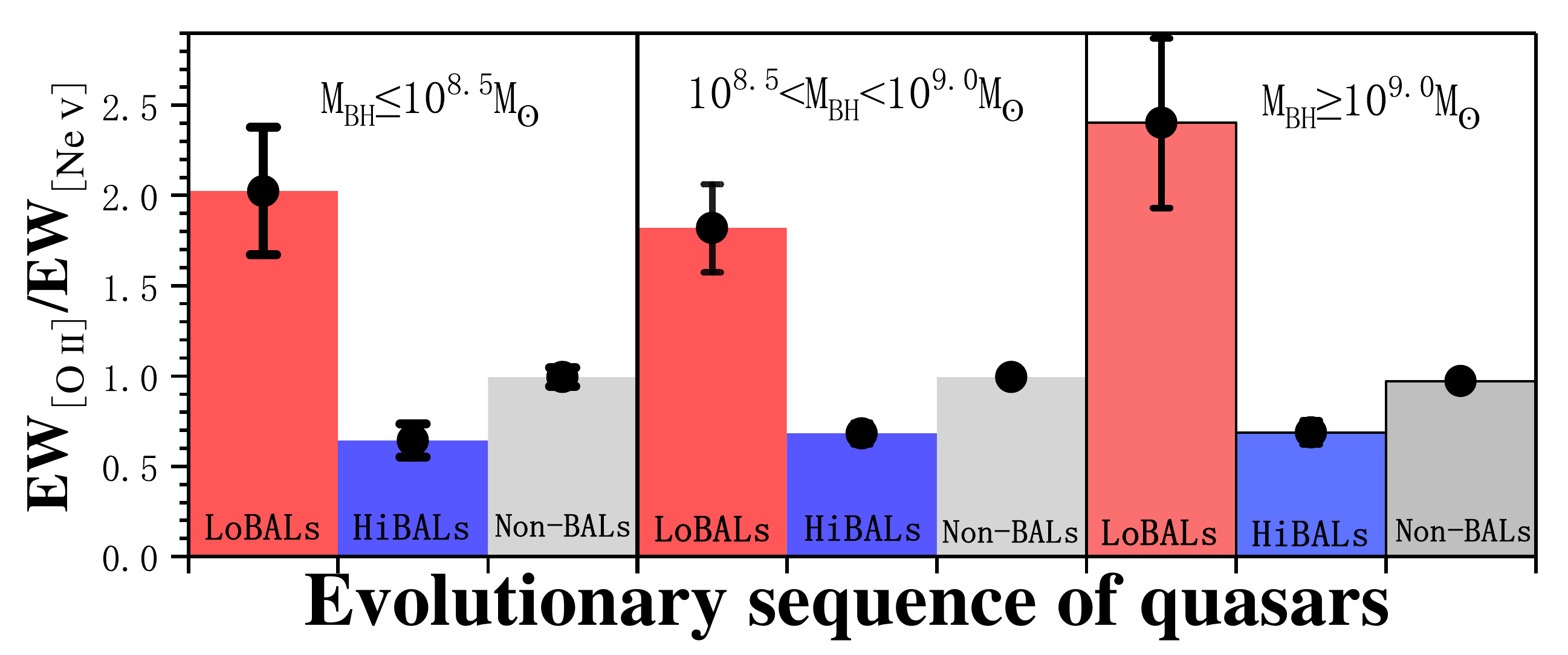}
\caption{\textbf{The variation of line ratio $R = {\rm EW_{\oii}/EW_{\nev}}$ as quasars evolve for different black hole masses.}
The line ratio $R$ significantly decreases and then increases as quasars evolve from 
LoBALs to HiBALs to non-BALs, for each of the three bins of black hole masses.
The vertical error bars mark the 1$\sigma$ uncertainty of the line ratio.}
\label{EDFig2}
\end{figure*}

\clearpage

\noindent \textbf{\huge Supplementary Information}
\addtocounter{figure}{-2}
\addtocounter{table}{-2}

%\begin{figure}
%\renewcommand{\figurename}{Extended Figure}
%\centering
%\includegraphics[width=0.85\textwidth]{figs/dist_z}
%\caption{Distributions of quasar redshifts. Black line is for the Non-BAL quasars, blue line is for the HiBAL quasars, and red line is for the LoBAL quasars.}
%\label{fig:dist_z}
%\end{figure}

\begin{table}[htbp]
\renewcommand{\tablename}{Supplementary Table}
\caption{The equivalent widths of \NEV\ and \OII\ of the quasars with different BALs ${\rm EW_{\civ}^{BAL}}$.} 
\centering \tabcolsep 3mm
\label{Tab:NEV_OII_ewbin}
\begin{tabular}{cccccc}
\hline\hline\noalign{\smallskip}
Samples   & ${\rm EW_{\nev}}$  &${\rm EW_{\oii}}$   & $R = \frac{{\rm EW_{\oii}}}{{\rm EW_{\nev}}}$  \\
          &\AA           &\AA           &                        \\
\hline\noalign{\smallskip}
${\rm EW^{\rm BAL}_{\civ}}\ge25$ \AA        & 0.582$\pm$0.0749  & 1.035$\pm$0.096   & 1.779$\pm$0.282   \\
$20\le {\rm EW^{\rm BAL}_{\civ}}<25$ \AA    & 1.539$\pm$0.085   & 1.066$\pm$0.133   & 0.693$\pm$0.095   \\
$15\le {\rm EW^{\rm BAL}_{\civ}}<20$ \AA    & 1.740$\pm$0.076   & 1.127$\pm$0.099   & 0.648$\pm$0.064   \\
$10\le {\rm EW^{\rm BAL}_{\civ}}<15$ \AA    & 2.170$\pm$0.112   & 1.296$\pm$0.127   & 0.597$\pm$0.066   \\
${\rm EW^{\rm BAL}_{\civ}}<10$ \AA         & 1.425$\pm$0.125   & 1.090$\pm$0.116   & 0.765$\pm$0.106   \\
\hline\hline\noalign{\smallskip}
\end{tabular}
\end{table}
%XX This figure looks ugly.  The left two panels look compressed along the X-axis compared to the right panels.  Panel (d) has too much empty white space on the bottom.  You should make the limits of the X-axis the same for all panels.
%
\begin{figure*}
\renewcommand{\figurename}{Supplementary Figure}
\centering
\includegraphics[width=7.5cm,height=6cm]{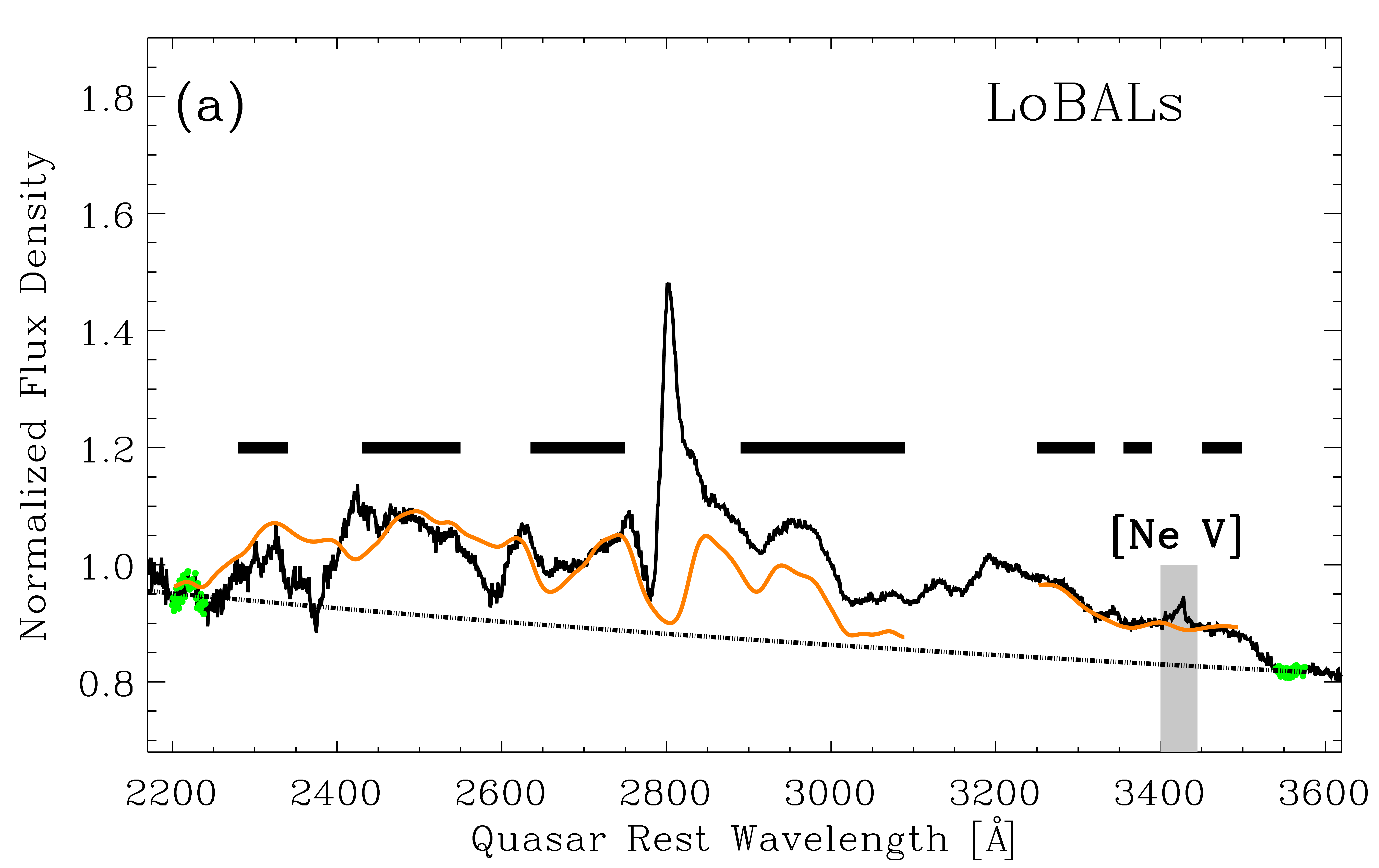}
\includegraphics[width=7.5cm,height=6cm]{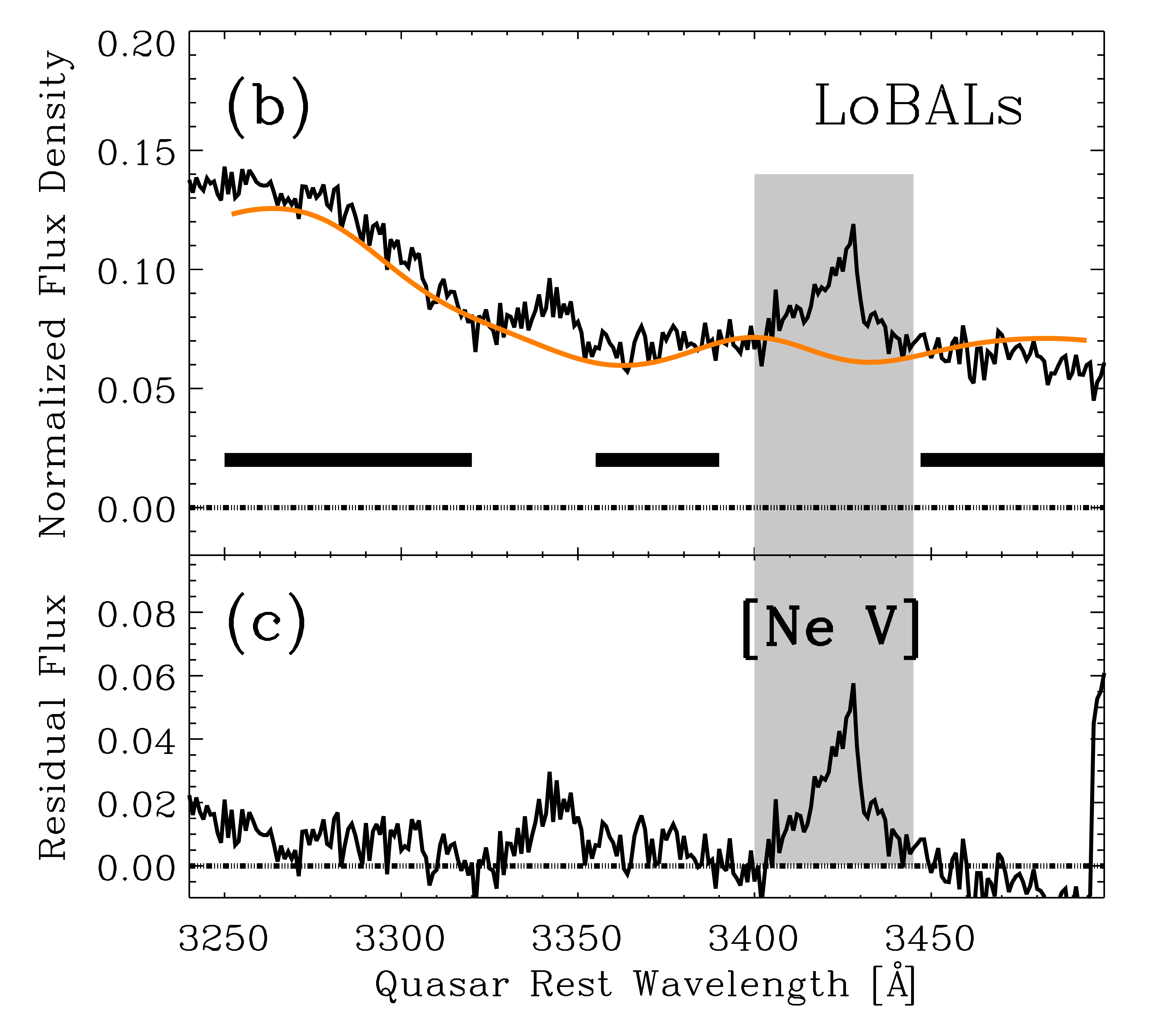}
\includegraphics[width=7.5cm,height=6cm]{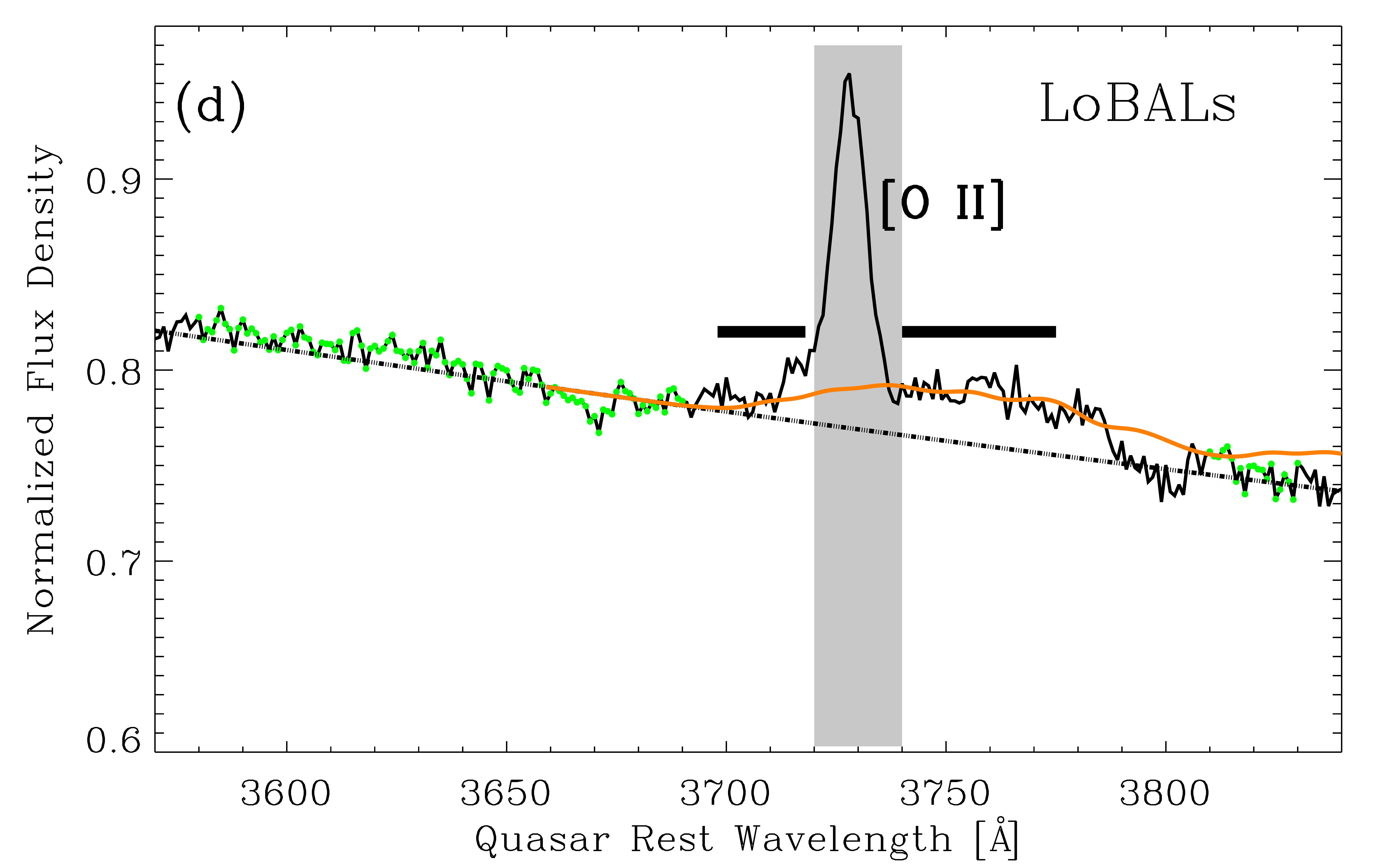}
\includegraphics[width=7.5cm,height=6cm]{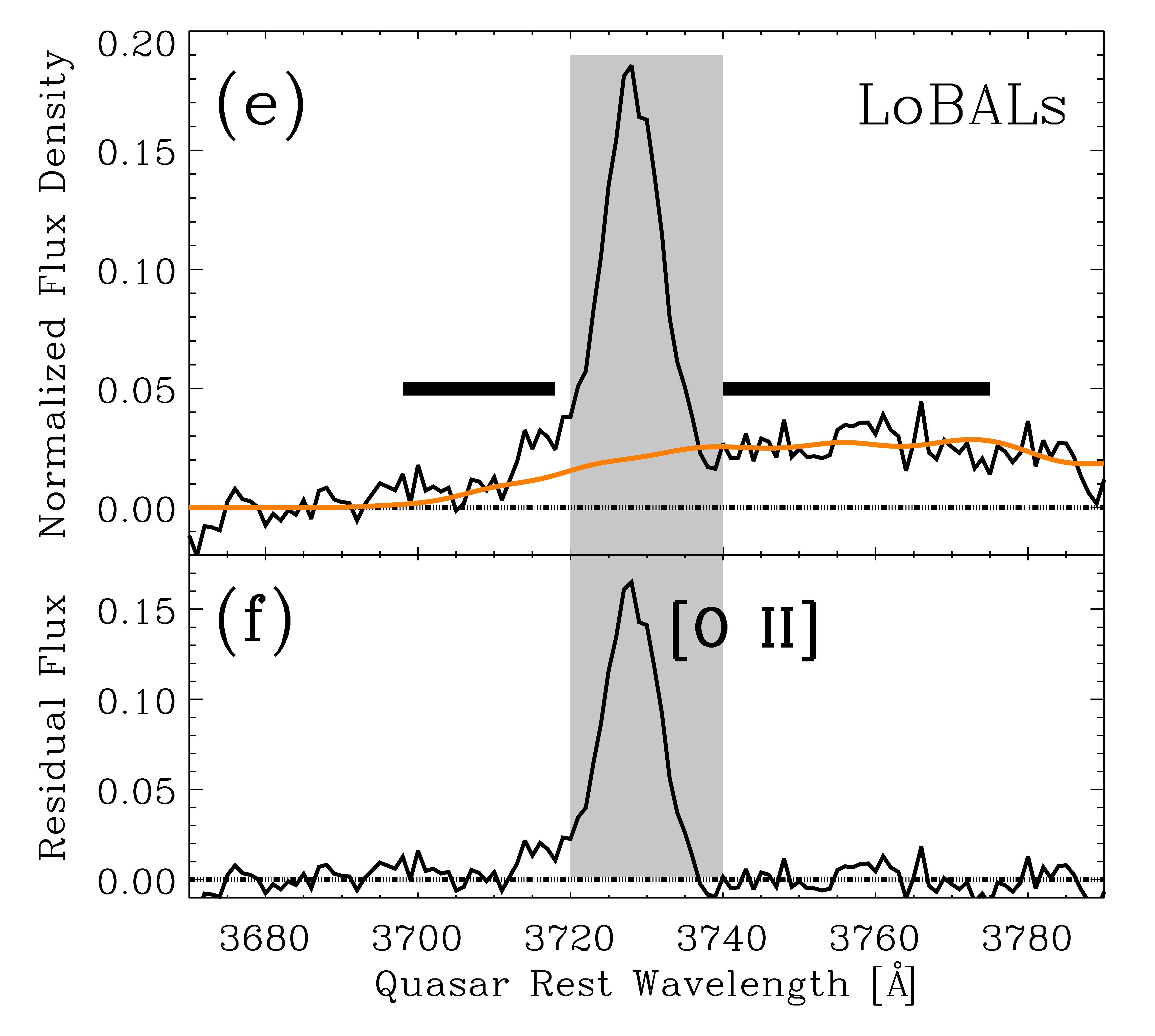}
\caption{\textbf{The median composite spectra of the LoBAL quasars.} Green points indicate the data used to fit the power-law continuum (black dash-dot lines). Black horizonal thick lines mark the spectral regions used to fit the iron template, and the orange solid lines represent the fitting results (power law plus iron template). Gray shaded regions identify the \oii\ and \nev\ lines. The left panels \textbf{a, d} show the global result from fitting the power-law continuum and iron template. In the right panels, \textbf{b, e} show the composite spectra after subtracting the best-fitting power law. \textbf{c, f} show the composite spectra after subtracting the best-fitting power law and iron template. }
\label{fig:composite_fitsLo}
\end{figure*}

%XX See same comments for Extended Figure 1
%
\begin{figure*}
\renewcommand{\figurename}{Supplementary Figure}
\centering
\includegraphics[width=7.5cm,height=6cm]{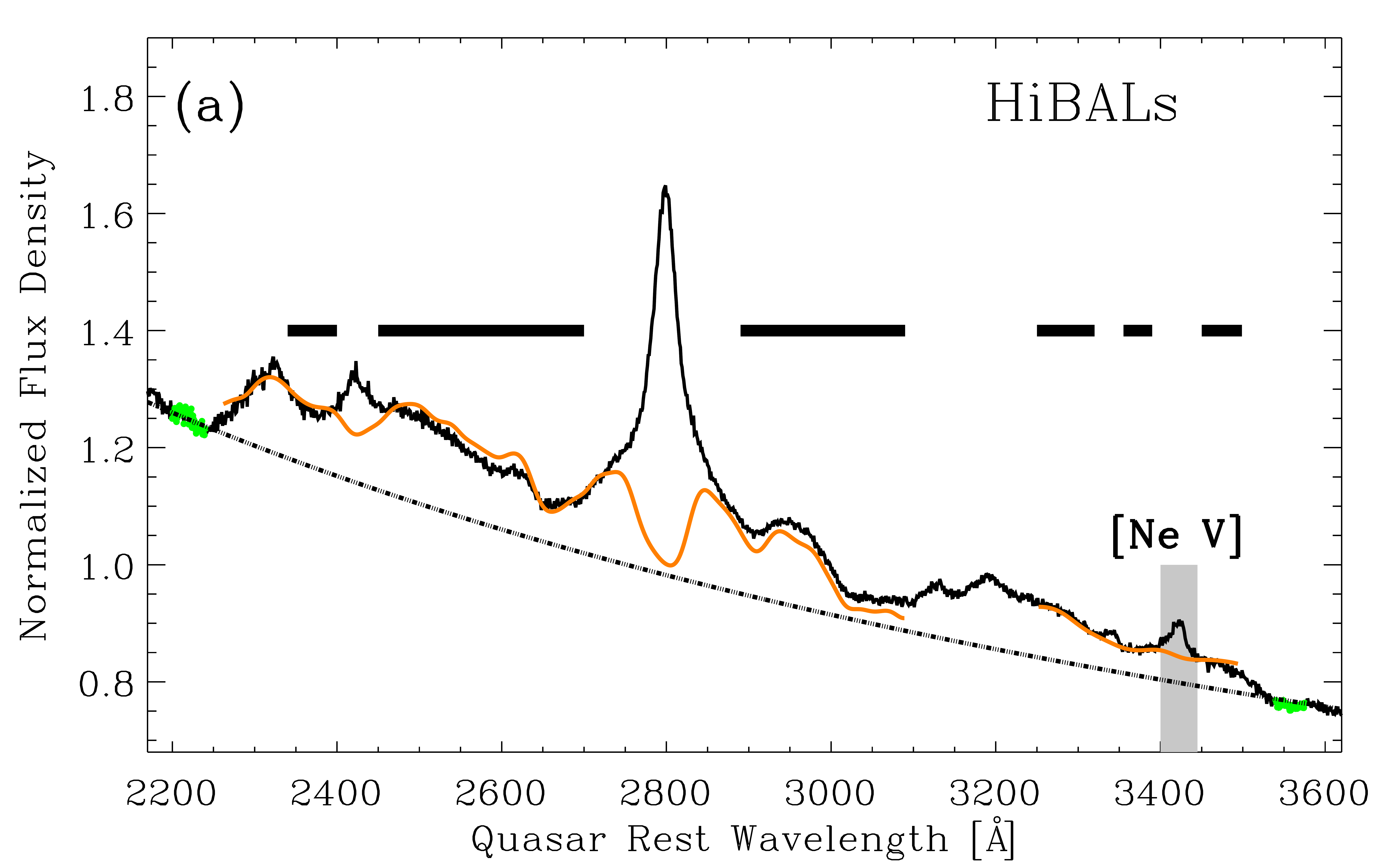}
\includegraphics[width=7.5cm,height=6cm]{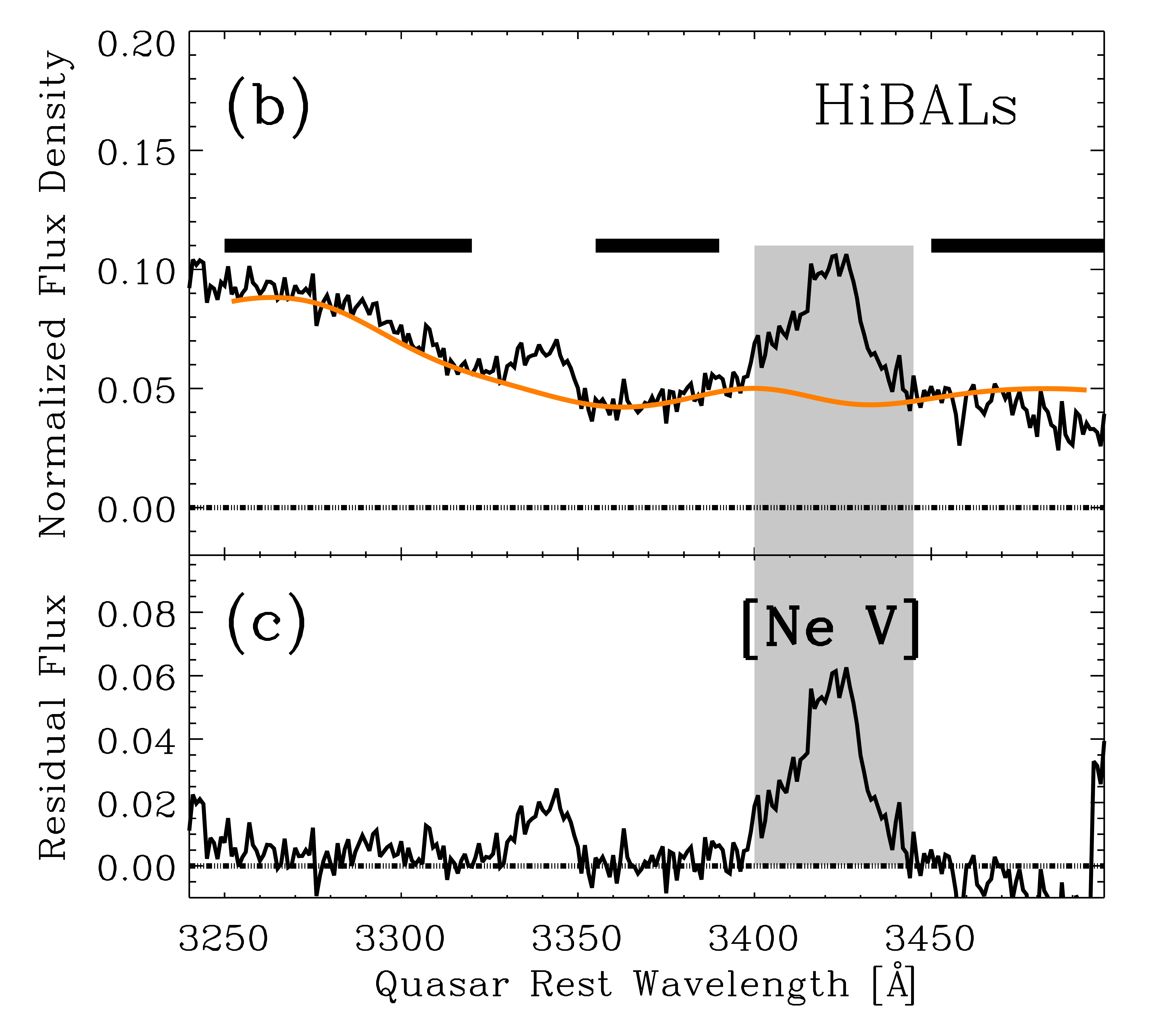}
\includegraphics[width=7.5cm,height=6cm]{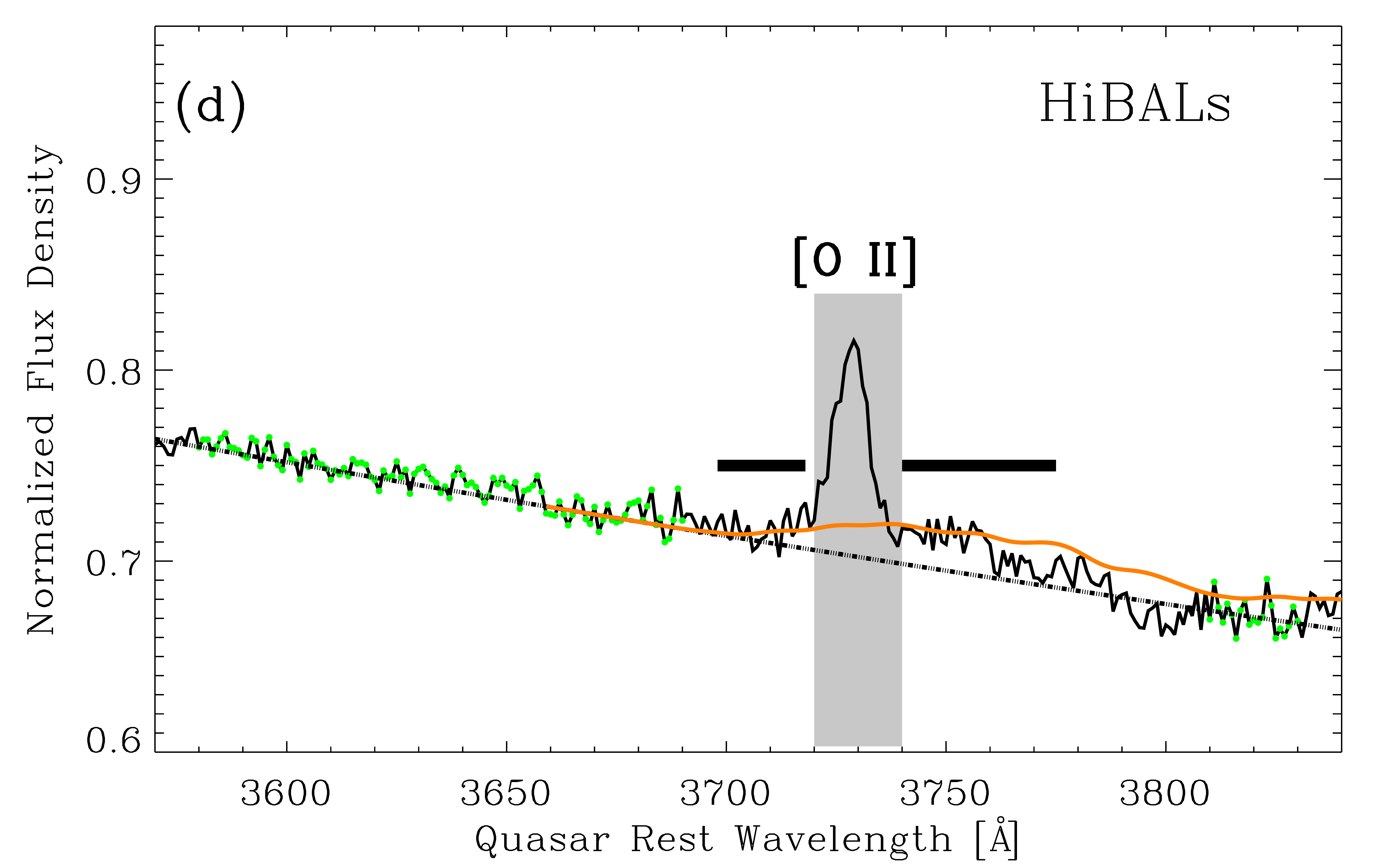}
\includegraphics[width=7.5cm,height=6cm]{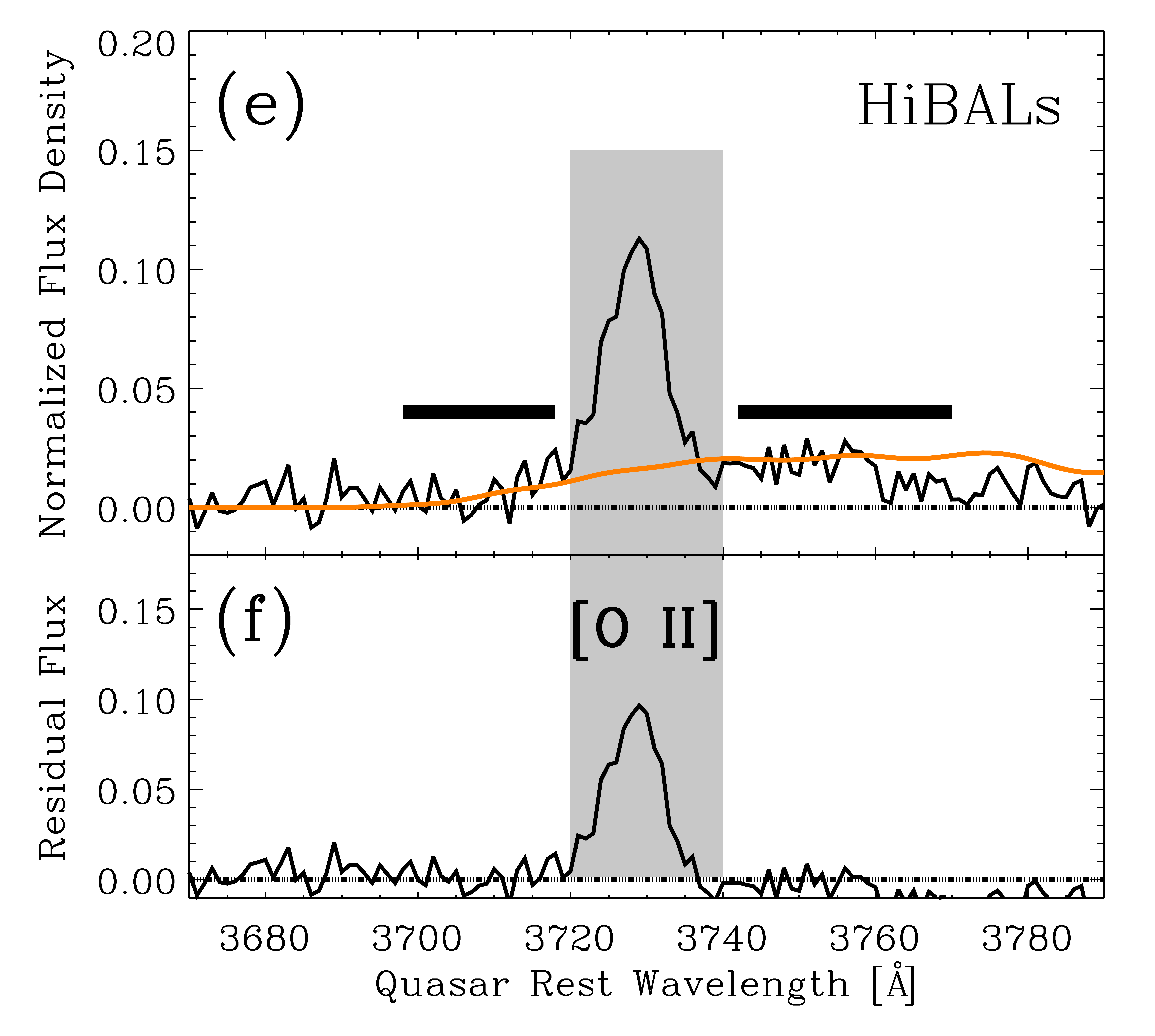}
\caption{\textbf{The median composite spectra of the HiBAL quasars.} Green points indicate the data used to fit the power-law continuum (black dash-dot lines). Black horizonal thick lines mark the spectral regions used to fit the iron template, and the orange solid lines represent the fitting results (power law plus iron template). Gray shaded regions identify the \oii\ and \nev\ lines. The left panels \textbf{a, d} show the global result from fitting the power-law continuum and iron template. In the right panels, \textbf{b, e} show the composite spectra after subtracting the best-fitting power law. \textbf{c, f} show the composite spectra after subtracting the best-fitting power law and iron template.}
\label{fig:composite_fitsHi}
\end{figure*}

%XX See same comments for Extended Figure 1
%
\begin{figure*}
\renewcommand{\figurename}{Supplementary Figure}
\centering
\includegraphics[width=7.5cm,height=6cm]{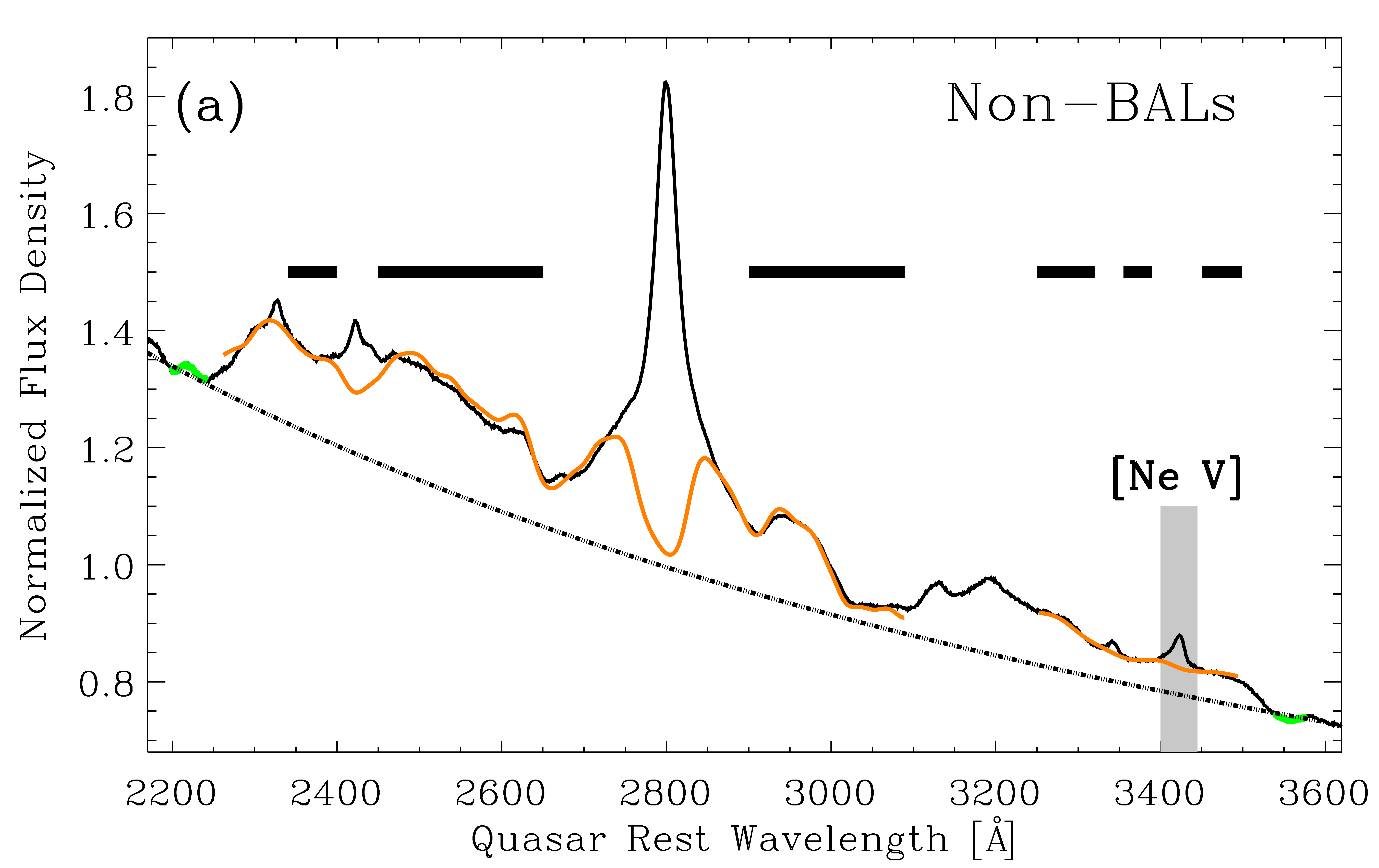}
\includegraphics[width=7.5cm,height=6cm]{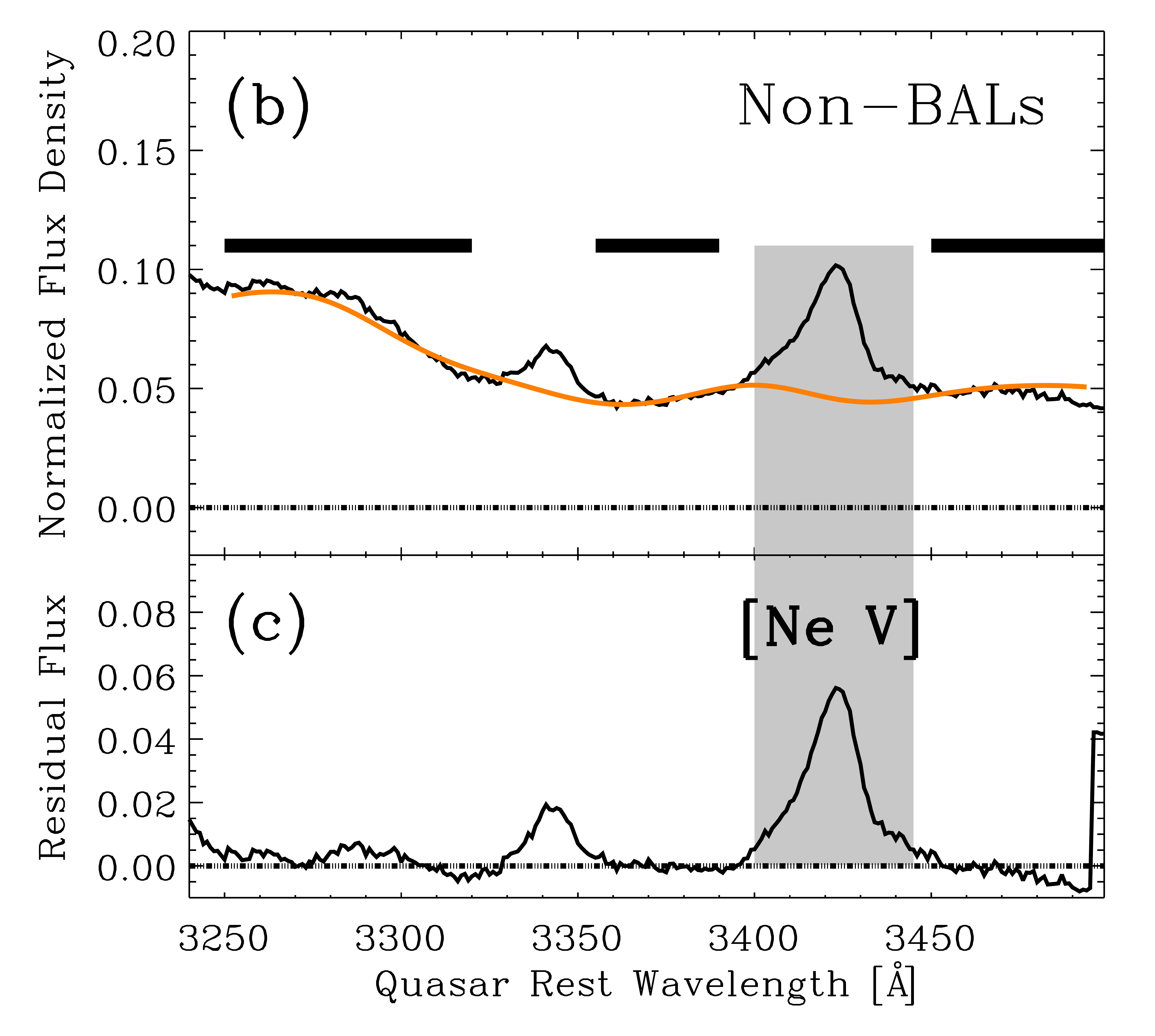}
\includegraphics[width=7.5cm,height=6cm]{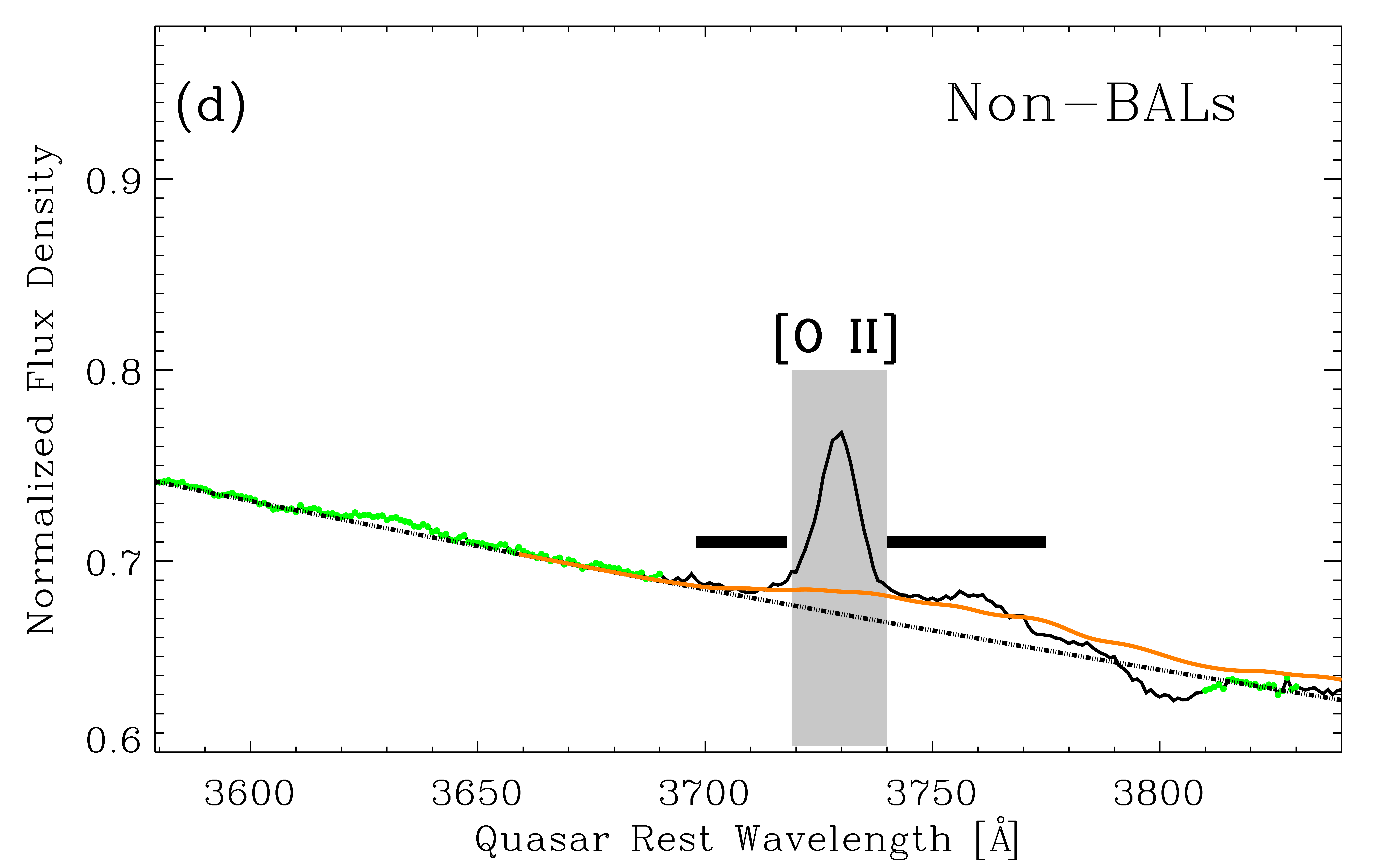}
\includegraphics[width=7.5cm,height=6cm]{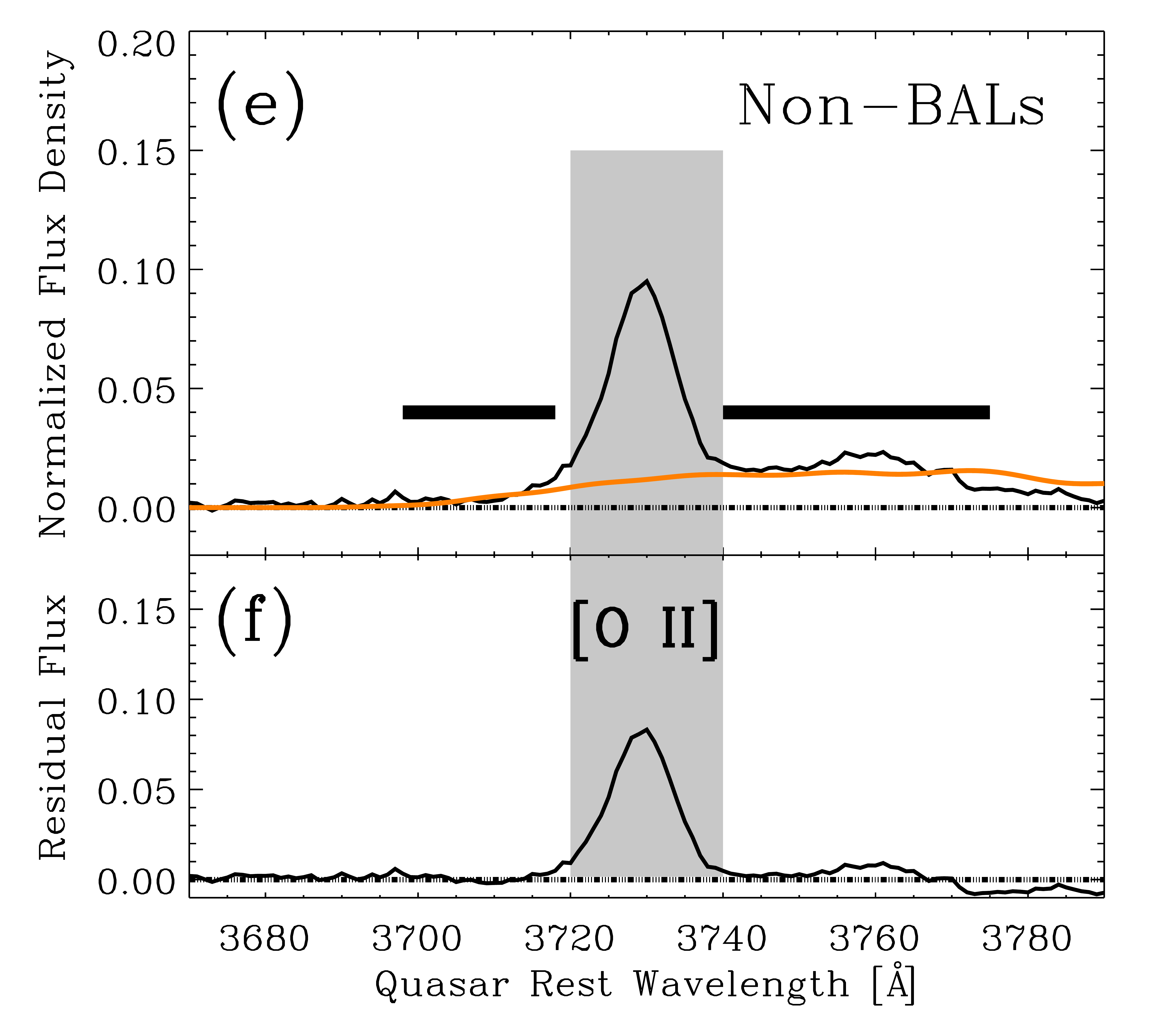}
\caption{\textbf{The median composite spectra of the non-BAL quasars.} Green points indicate the data used to fit the power-law continuum (black dash-dot lines). Black horizonal thick lines mark the spectral regions used to fit the iron template, and the orange solid lines represent the fitting results (power law plus iron template). Gray shaded regions identify the \oii\ and \nev\ lines. The left panels \textbf{a, d} show the global result from fitting the power-law continuum and iron template. In the right panels, \textbf{b, e} show the composite spectra after subtracting the best-fitting power law. \textbf{c, f} show the composite spectra after subtracting the best-fitting power law and iron template.}
\label{fig:composite_fitsNo}
\end{figure*}

\begin{figure*}
\renewcommand{\figurename}{Supplementary Figure}
\centering
\includegraphics[width=0.85\textwidth]{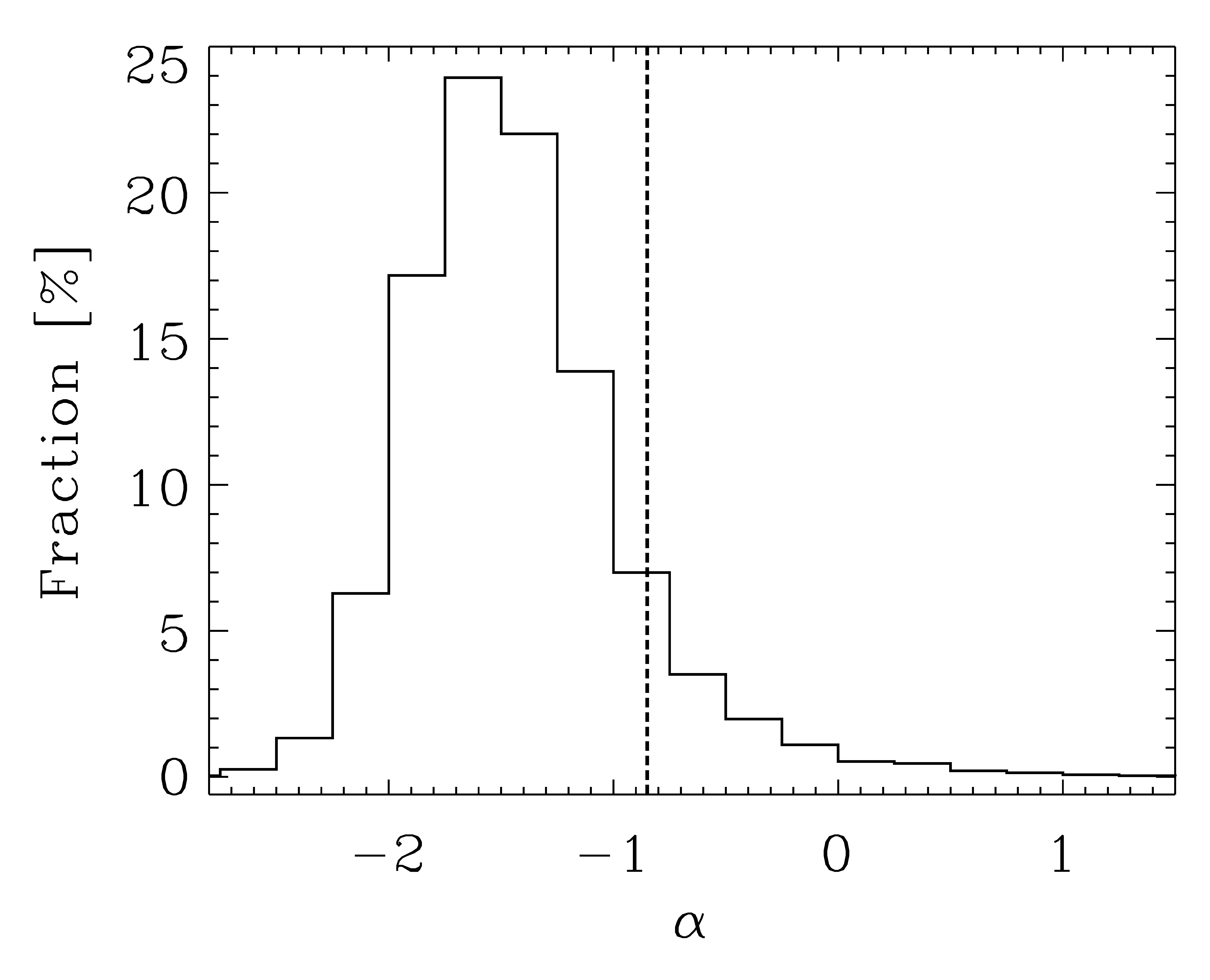}
\caption{\textbf{The distribution of spectral index of the non-BAL quasars.} 
The vertical dashed line indicates $\alpha=-0.85$; $\sim$10\% of sources have $\alpha>-0.85$. }
\label{fig:slope}
\end{figure*}

\begin{figure*}
\renewcommand{\figurename}{Supplementary Figure}
\centering
\includegraphics[width=0.49\textwidth]{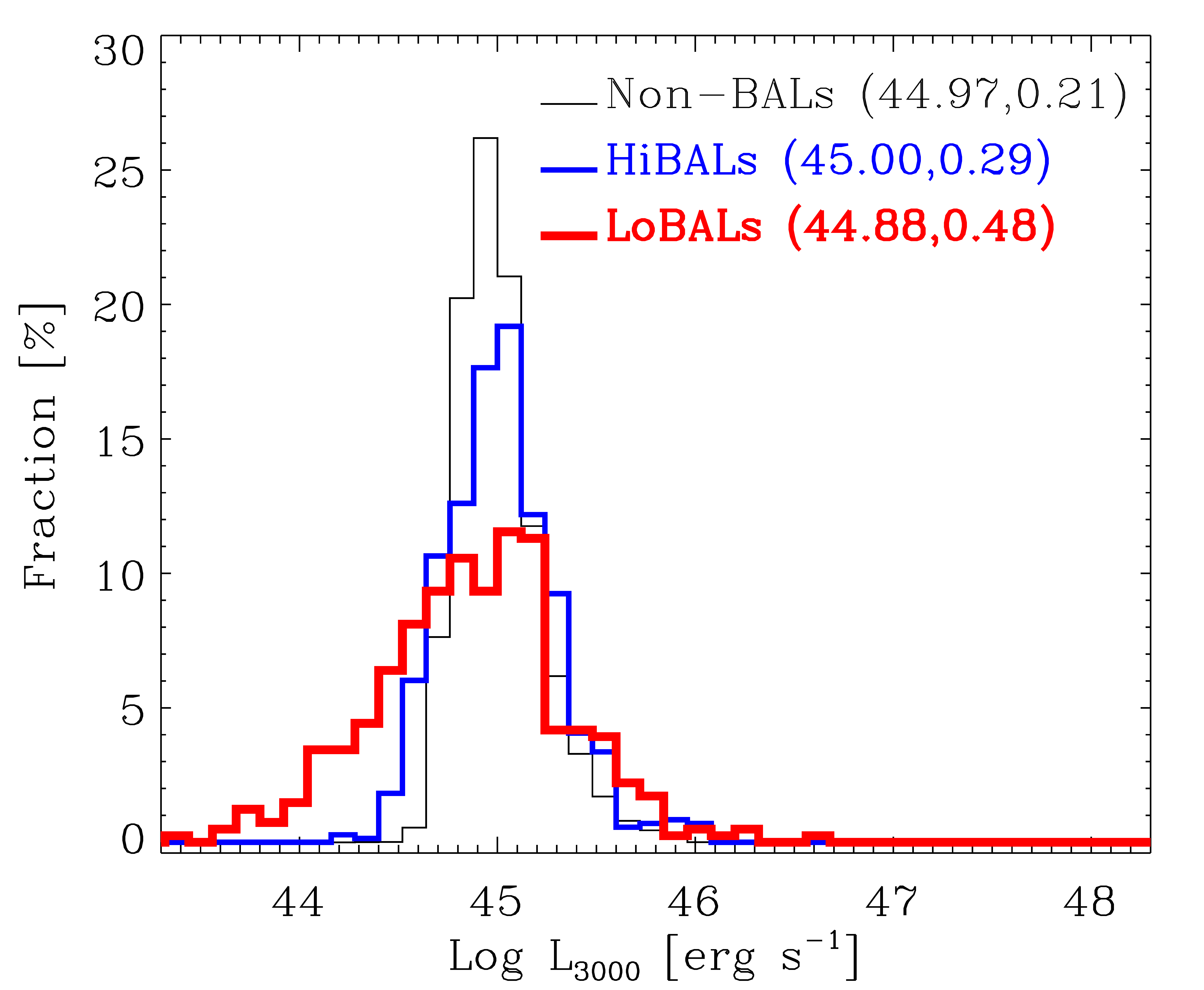}
\includegraphics[width=0.49\textwidth]{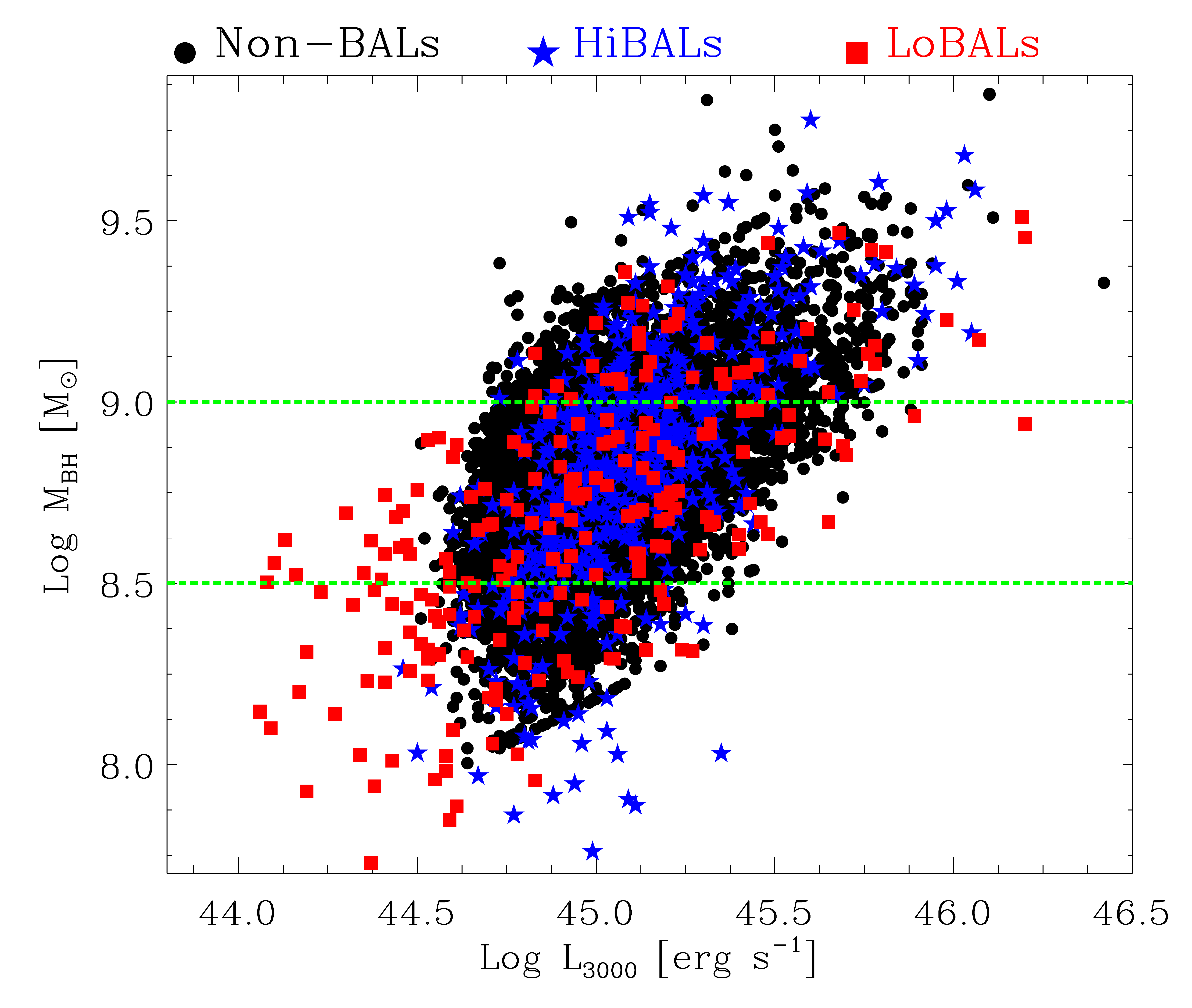}
\caption{\textbf{The primary parameters distributions of quasar samples.} Left panel: The distribution of 3000 \AA\ luminosity for the three classes of quasars. The legend gives the median values and standard deviations of the distributions. Right panel: The distribution of 3000 \AA\ luminosity versus black hole mass. Green dashed lines indicate the levels of $M_{\rm BH}=10^{8.5}$ and $10^{9.0}~M_{\odot}$. }
\label{fig:compare_L}
\end{figure*}

\begin{figure*}
\renewcommand{\figurename}{Supplementary Figure}
\centering
%\vspace{-1ex}
\includegraphics[width=0.81\textwidth]{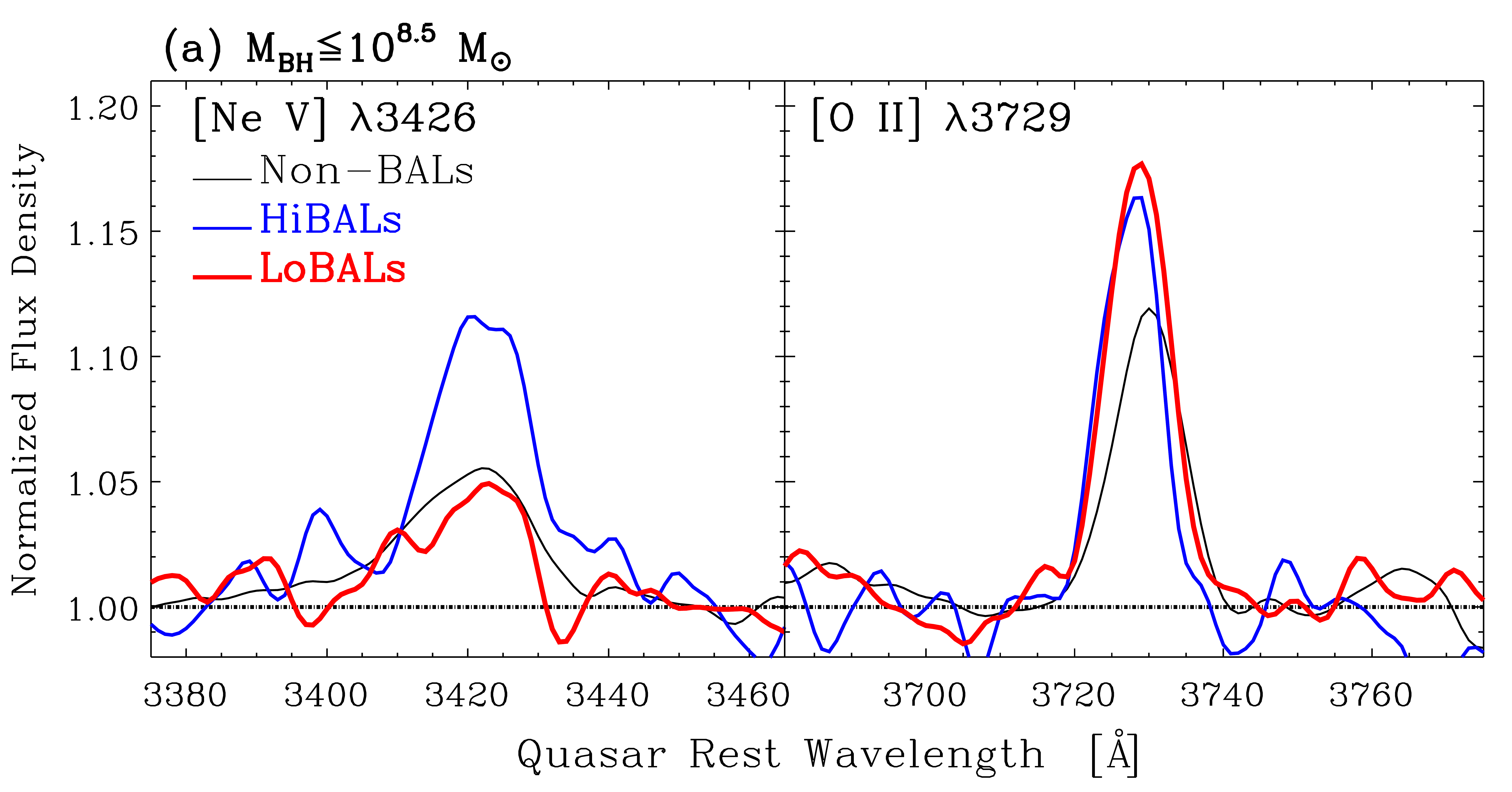}
\vspace{-1ex}
\includegraphics[width=0.81\textwidth]{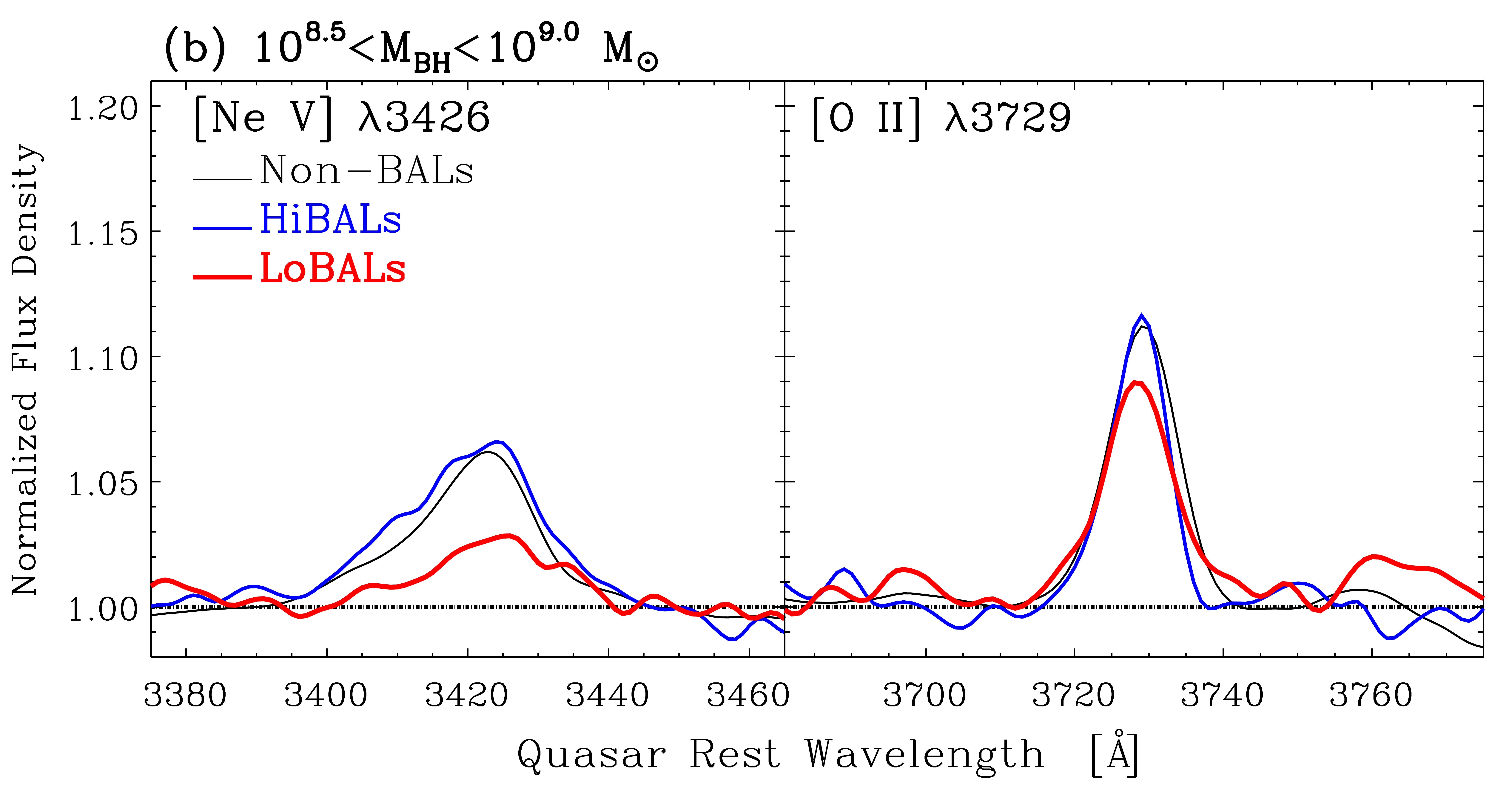}
\vspace{-2ex}
\includegraphics[width=0.81\textwidth]{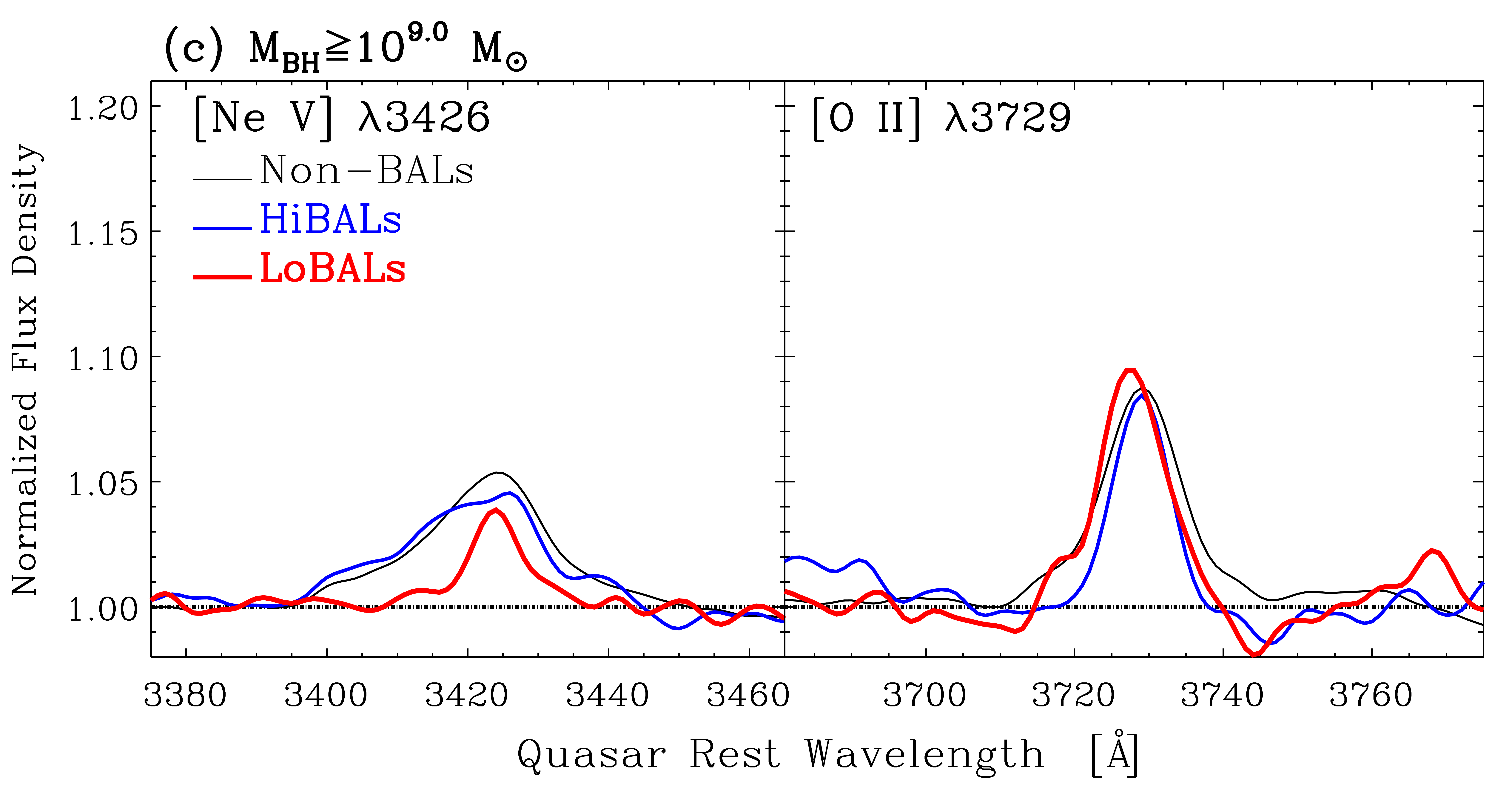}
\caption{\textbf{The \nev\ and \oii\ emission lines in the median composite spectra of quasars with different black hole masses.}.
 \textbf{a} is for $M_{\rm BH}\le10^{8.5}~M_{\odot}$, \textbf{b} is for $10^{8.5} < M_{\rm BH} < 10^{9.0}~M_{\odot}$, and \textbf{c} 
 is for $M_{\rm BH} \ge 10^{9.0}~M_{\odot}$.}
\label{fig:composite_fits_bin_lombh}
\end{figure*}

\begin{figure*}
\renewcommand{\figurename}{Supplementary Figure}
\centering
\includegraphics[width=1.0\textwidth]{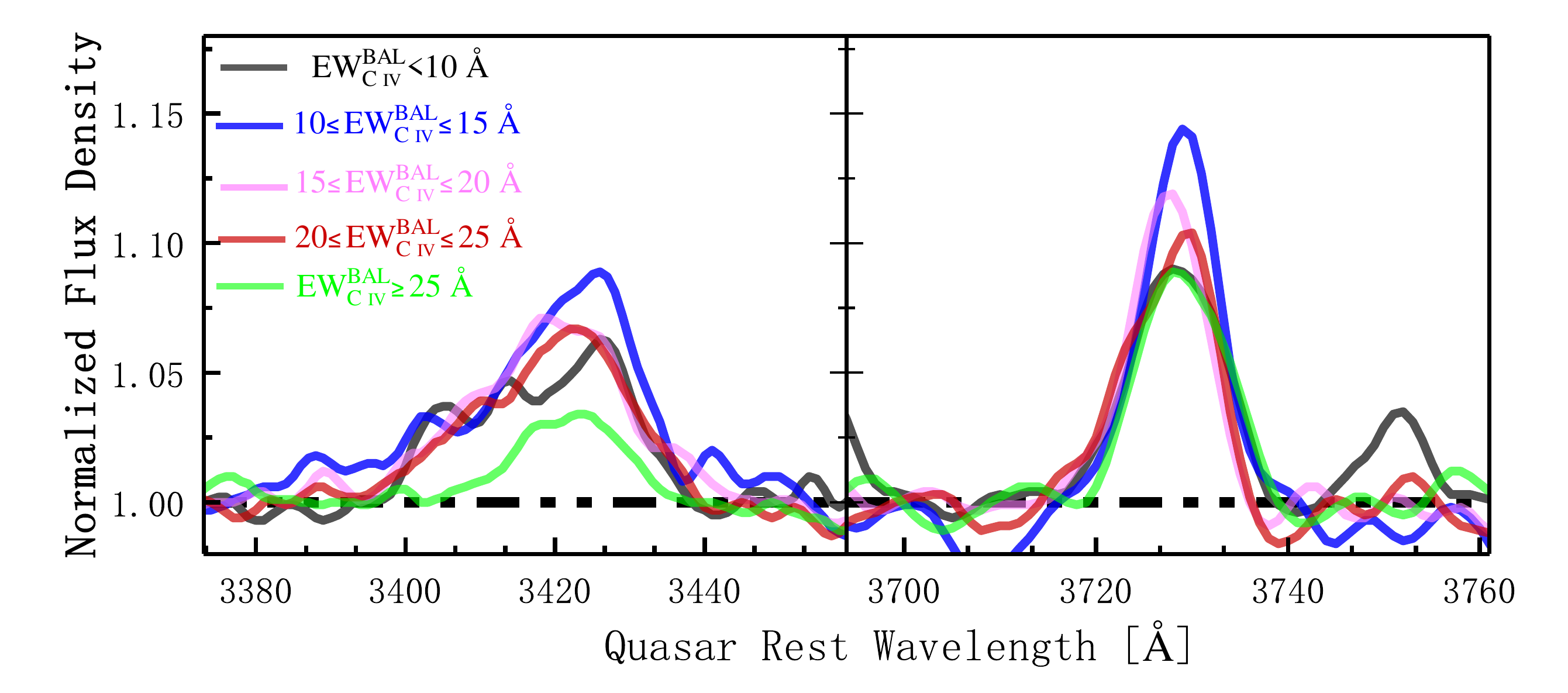}
\caption{\textbf{The \nev\ and \oii\ emission lines in the median composite spectra of quasars with different ${\rm EW_{\civ}^{BAL}}$.}
Left panel is for \nev\ while right panel is for \oii . }
\label{fig:composite_fits_bin}
\end{figure*}
\end{document}